\documentclass[usenatbib,useAMS,usegraphicx]{mn2e}

\usepackage{times}

\usepackage{xspace}
\usepackage[fleqn]{amsmath}
\usepackage[rightcaption]{sidecap}
\usepackage{comment}
\usepackage{mathtools}

\output\expandafter{\the\output\floatfix}

\makeatletter
\def\floatfix{%
\expandafter\ifx\csname r@x@one\endcsname\relax
\else
\ifnum\c@page=\numexpr\expandafter\expandafter\expandafter
              \@secondoftwo\csname r@x@one\endcsname-1\relax
\aftergroup\figone
\fi
\fi}
\makeatother

\providecommand{\eg    }{e.g.\xspace}%
\providecommand{\ie    }{i.e.\xspace}
\providecommand{\mrk   }{Mrk~421\xspace}%
\providecommand{\xray  }{X-ray\xspace}%
\providecommand{\gray  }{$\gamma$-ray\xspace}%
\providecommand{\grays }{$\gamma$-rays\xspace}%

\usepackage{colordvi}
\newcommand\be{\begin{equation}}
\newcommand\ee{\end{equation}}
\newcommand\bal{\begin{align}}
\newcommand\eal{\end{align}}

\title[Particle diffusion and localized acceleration]
{Particle diffusion and localized acceleration in inhomogeneous AGN jets -
Part I: Steady-state spectra}
\author[X. Chen et al.]{%
\parbox{\textwidth}{Xuhui~Chen$^{1,2}$\thanks{chenxuhui.phys@gmail.com},
Martin~Pohl$^{1,2}$,
Markus~B\"ottcher$^{3,4}$
}\vspace{0.1cm}\\
$^1$ Institute of Physics and Astronomy, University of Potsdam, 14476 Potsdam-Golm, Germany\\
$^2$ DESY, Platanenallee 6, 15738 Zeuthen, Germany\\
$^3$ Centre for Space Research, North-West University, Potchefstroom 2520, South Africa\\
$^4$ Astrophysical Institute, Department of Physics and Astronomy, Ohio University, Athens, OH 45701, USA\\
      }

\begin{document}
%\date{}

\maketitle

\label{firstpage}

\begin{abstract}
We study the acceleration, transport, and emission of particles in relativistic jets. 
Localized stochastic particle acceleration, spatial diffusion, 
and synchrotron as well as synchrotron self-Compton emission are considered in a leptonic model.
To account for inhomogeneity, 
we use a 2D axi-symmetric cylindrical geometry for both relativistic electrons and magnetic field.
In this first phase of our work, we focus on steady-state spectra that develop from a time-dependent model.
We demonstrate that small isolated acceleration region in a much larger emission volume are sufficient to 
accelerate particles to high energy. Diffusive escape from these small regions provides a natural explanation 
for the spectral form of the jet emission.
The location of the acceleration regions within the jet
is found to affect the cooling break of the spectrum in this diffusive model.
Diffusion-caused energy-dependent inhomogeneity in the jets predicts that 
the SSC spectrum is harder than the synchrotron spectrum. 
There can also be a spectral hardening towards the high-energy section of the synchrotron spectrum,
if particle escape is relatively slow. 
These two spectral hardening effects indicate that the jet inhomogeneity might be a
natural explanation for the unexpected hard \gray spectra observed in some blazars.

\end{abstract}
\section{Introduction}
\label{intro}
The inner parts of relativistic jets in Active Galactic Nuclei (AGNs) are known to emit radiation
in every energy band we can observe. The actual size and location of the emission region, 
\eg those in blazars, are still under debate
\citep{gg_tavecchio:2009:canonical_blazars, marscher_2013:granada_review}. 
Their size and distance make 
them challenging to resolve with our current
% fall below the current resolution of 
imaging capability, 
except maybe a few cases where mm-VLBI observations are paving the way to resolve the base of the jet
\citep{lu_2013:3c279_mmvlbi.772.13, doeleman_2012:m87_mmvlbi.338.355}. 
For this reason, many theoretical efforts
concerning AGN jets assume homogeneous emission region as the source of the multiwavelength emission
\citep[\eg][]{dermer_etal:2009:analysis_of_FSRQ}.

However, increasing temporal coverage of multiwavelength data and modeling results begin to suggest that 
single-zone
homogeneous models are not sufficient in describing the complex phenomena. 
The observation
that the blazars exhibit variability as fast as 3$\sim$5 minutes
\citep{albert_etal:2007:mrk501_fast_flares, aharonian_etal:2007:pks2155_exceptional_flare},
and the detection of \gray above 100\,GeV from several flat-spectrum radio quasars (FSRQs)
without signature of $\gamma-\gamma$ absorption
by soft photons in the broad-line region 
\citep{albert_etal:2008:3C279_magic_detection, hess_2013:1510, 
alksic_2011:1222_magic_detection:730.8},
indicate that the \gray emission region is
extremely small, and at the same time located at parsecs away from the central AGN engine.
This would require an unusually small angle of collimation, if the emission region covers 
the entire cross-section of the jet.
One resolution to this conflict is the hypothesis that the larger jet contains small high-energy regions,
presumablely resulting from turbulence that is generated locally, far way from the central black hole.
Apparently, single-zone homogeneous models are not adequate to describe these scenarios.
\citep[See][for an example of such turbulent blazar emission model.]{marscher_2014:turb_model.780.87}

In the picture considering small-scale structures, fast escape of particles
means that, the highest-energy
particles could have already cooled before they can travel far, while particles with lower energy 
still survive and occupy significant larger regions.
This consideration suggests that to account for the multiwaveband radiation signals of AGN jets,
one must consider inhomogeneous models spanning certain scale ranges to cover both the acceleration region
and the region with the escaped particles.
Various efforts have been made to model inhomogeneous jets 
\citep[\eg][]{gmt85, sokolov_marscher_mchardy:2004:ssc,sokolov_marscher:2005:ec,
graff_etal:2008:pipe}, 
although usually the details of the particle acceleration were not considered.
Simplified approaches have been adopted to treat the acceleration region 
and the emission region separately \citep{kirk_rieger_mastichiadis:1998}, although the emission from the
acceleration region is not considered in their case.
Recently, \citet{richter:2015:resolvedmodel.61.102} built a 
one-dimensional spatially-resolved model that accounts for 
the particle acceleration process, but the important light-travel-time effects (LTTEs) are not considered.
Their geometry is suited for the laterally homogeneous shock structure, but not suitable for the study
of 2D/3D small-scale structures such as turbulent acceleration regions.

\begin{figure}
\includegraphics[width=0.99\linewidth]{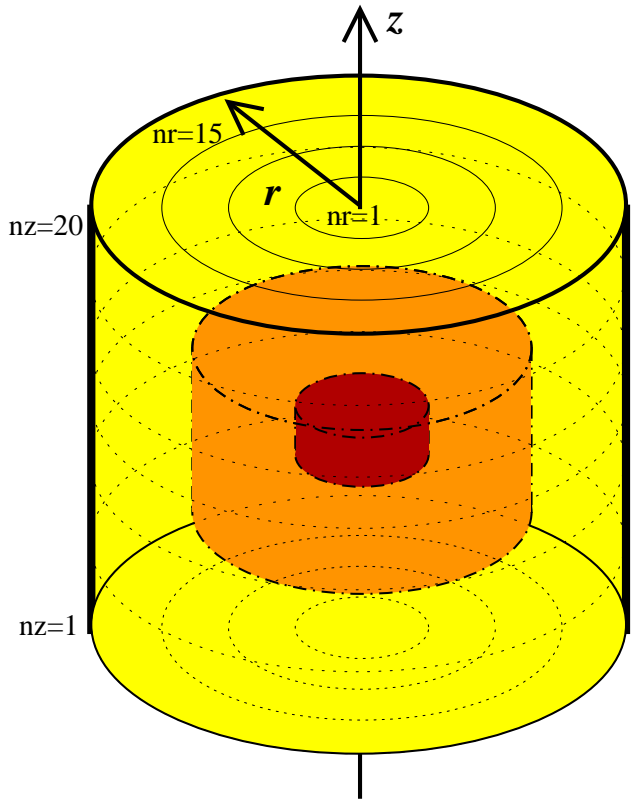}
\caption{Sketch of the 3D geometry of the particle diffusion and localized acceleration in the
axisymmetric cylinder. The red region represents the acceleration region
where the acceleration is causing the particles to have the highest energy density. 
The spatial diffusion causes its surrounding regions to be still relatively energetic (orange zone), 
while in the yellow zone the particles have already cooled significantly.
The actual particle distribution is shown in more detail in Fig.\,\ref{fig:refl_ct_4emaps} and other figures as 2D maps.
}
\label{fig:geometry}
\end{figure}

\begin{figure}
\centering
\includegraphics[width=0.99\linewidth]{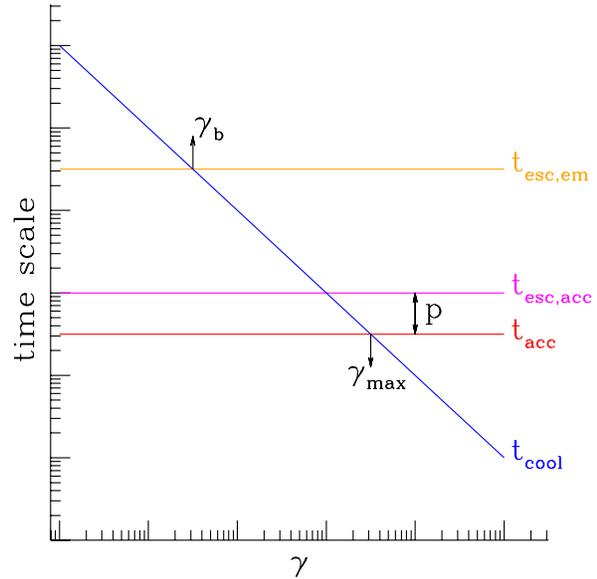}
\caption{A sketch of the relationship between various time scales related to particle acceleration,
cooling, and escape. 
}
\label{fig:tscale} 
\end{figure}

The modeling of blazar SEDs usually requires very fast particle escape, 
considering the electron spectrum form required to match the observation.
The required escape time scale is usually not much longer than the light-crossing time of 
the blazar emission region \citep{katarzynski_etal:2006:stochastic, chen_2014:bamplify:441.2188C}.
For this fast escape to be physically feasible, this escape should refer to escape from 
the accelerator, probably some smaller-scale structures \citep{giannios_2013:plasmoid.431.355} 
within the emission region.
Particles experience cooling and diffusion, but no acceleration outside of these regions.
Throughout this paper, we will call the entire region of the jet contributing to 
the blazar radiation `emission region'. The smaller sub-region where particle acceleration takes place is 
referred to as `acceleration region', while the rest of the `emission region' 
takes the name `diffusion region'. 

Particle acceleration mechanisms that predict localized acceleration confined in small regions
include magnetic reconnection \citep{guo_2014:reconnection_pic,sironi_2014:reconnection.783.21}, which can be triggered through turbulence
\citep{zhang_yan_2011:icmart.726.90}, as well as various acceleration mechanisms at the shock front \citep{blandford_eichler:1987, sironi_2011:shock_acc.726.75}. 
The thin but extended structure of the shocks 
\citep[\eg, internal, external, or standing shocks,][]{spada_etal:2001:internal_shocks, 
kirk_rieger_mastichiadis:1998, nalewajko_2009:reconfinement_shock.392.1205} means that they can facilitate fast particle escape, 
but 
in order to explain the fast variability in blazar emission, 
internal shocks produced very close to the jet base \citep{rachen_2010:discontinuous:1006.5364}, or
shocks associated with mini-jets \citep{giannios_etal:2009:jets_in_a_jet} or, again, small scale turbulent structures \citep{marscher_2014:turb_model.780.87}, would be required.

Except the consideration of emission region structure and particle acceleration, 
another major focus of relativistic jet models has been the radiative mechanism.
The SEDs of blazars usually consist of two components, with the first peaking between infrared to 
\xray frequencies, while the second peaking between \xray to \gray energies 
\citep{ulrich_maraschi_urry:1997:review, fossati_etal:1998:sequence}.
Both hadronic and leptonic models have been frequently discussed, and have been
successfully applied to blazars in most cases
\citep{boettcher_2013.leptonic_hadronic.768.54}.
The two kinds of models agree in explaining the low-frequency (below ultra-violet or \xray) 
component of the blazar emission as electron synchrotron emission, but differ in their interpretation
of the origin of the high energy (above \xray) component.
In the hadronic models protons are responsible for the high energy radiation through processes such as
proton synchrotron emission \citep{aharonian:2000:proton_synchrotron, mucke_protheroe:2001:proton_synchrotron},
p-p pion production \citep{pohl_2000:blastwave:354.395}, or p-$\gamma$ pion production \citep{mannheim_biermann:1992} with subsequent synchrotron emission of pion decay products 
\citep{mannheim:1993:proton_blazar,rachen_correlated_x_tev, mucke_etal:2003:proton_synchrotron}.
The leptonic models on the other hand assume that the electrons, and possibly also positrons, 
in addition to providing the low-frequency emission through synchrotron, are also responsible for
the high energy emission through inverse Compton (IC) scattering \citep[\eg][]{mgc92_3c279}.
Depending on whether the seed photons of these scattering are the synchrotron photons the leptons themselves
produced, or photons with origin external to the jet, the leptonic models can be further classified into
synchrotron self-Compton (SSC) models and external Compton (EC) models. The EC models can then differ from 
each other based on the various possible sources of external seed photons, 
such as the accretion disc \citep{dermer_etal92}, 
the broad line region \citep{gg_madau:1996}, 
or the dusty torus \citep{sikora_etal:2009:constraining_emission_models}. 
The complexity of the SSC models come from
the mathematical treatment of the nonlinear cooling of electrons in the SSC process
\citep{zacharias_2013:sync_lightcurve_ssc.777.109, zacharias_2014:ic_lightcurve_ssc.443.3001}, 
which is further complicated by the light retardation of the synchrotron photons 
\citep[part of LTTEs, see the discussion by][]{sokolov_marscher_mchardy:2004:ssc}.
Traditionally the SSC models are usually associated with BL Lac objects 
while the EC models are usually associated with FSRQs.
This is because, by definition, external emission lines are readily seen in FSRQs, but not in BL Lacs
\citep{ghisellini_etal:1998:seds}.
But whether this distinction in radiation mechanism is real or not, remains to be an open question
\citep{chen_etal:2012:424.789}.

In order to study inhomogeneous jets, \citet{chen_etal:2011:multizone_code_mrk421} have built a 
2D leptonic model that takes into account all the LTTEs, 
including the external ones that cause delayed observation of further-away cells,
and the internal ones that cause delayed arrivals of synchrotron photons in 
the SSC scattering. 
The model has been applied to cases
where the inhomogeneity is caused by plasma crossing a standing perturbation. In those cases the inhomogeneity
is mostly along the longitude of the jet, while the radial structure remains largely homogeneous.
Direct particle exchange between cells is also neglected, based on the fact that the Larmor radius of 
the electrons is sufficiently small, and the assumption that the magnetic field is highly tangled.
However, the nature of particle diffusion is also dependent on the turbulence property of the magnetic field,
which is poorly known. Under certain circumstances, the diffusion between cells can be very important.

In this work we extend the model of \citet{chen_etal:2011:multizone_code_mrk421} 
by implementing particle diffusion between cells, as one mechanism for realistic particle escape.
Combined with our direct handling of the particle acceleration using the Fokker-Planck equation,
we investigate both spatial and momentum diffusion of particles at the same time.
For the first time, our modeling of the particle evolution and emission 
encompasses both particles inside the accelerator and those already escaped from the accelerator.
A sketch of acceleration and emission regions is shown in Fig.\,\ref{fig:geometry}.
Although our model is a time dependent one, in this paper we focus on what kind of steady state spectra
emerge from the time dependent solution, and how. The
flare-related variations will be the topic of discussion in a forthcoming paper.

As a simplification that permits understanding some, but not all of the physics that is
captured in the 2D
model, we will first introduce an semi-analytical two-zone model in \S \ref{2zone}.
The methods used in the 2D
model will be described in \S \ref{model}, followed by the simulation 
results in \S \ref{results}. Discussion and conclusion can be found in 
\S \ref{discussion} and \S \ref{conclusion}.

Throughout this paper, we will use non-primed notations for the quantities in the jet frame, and primed ones
for those in the observer's frame. Subscripts `em',`acc' and `dif' are used to denote parameters for
emission, acceleration and diffusion regions respectively.

\section{A two-zone model}
\label{2zone}
We first discuss the particle and emission spectra resulting from a semi-analytical two-zone model,
which treats the acceleration and diffusion regions as two separate model zones.
In this two-zone model it is assumed that particles are injected and accelerated in a small 
spherical acceleration zone.
Those particles escape, and are subsequently injected into a much larger diffusion zone
that surrounds the acceleration zone. There is no particle acceleration in the diffusion zone,
but radiative cooling and further particle escape do play a role. In the two-zone model we only account for
the synchrotron cooling, while IC cooling is not considered.
We calculate analytically, with the help of numerical integrations,
the electron spectrum of particles in both the acceleration
and the diffusion zones. Then we estimate the synchrotron and SSC emission from both zones. We take into
account the synchrotron seed photons from both zones when calculating the SSC emission, under the
spherical geometry where the acceleration zone sits in the center of the diffusion zone.
The diffusion zone approximately generates a synchrotron photon energy density of 
$3L_\mathrm{s,dif}(\epsilon)/4\pi R_\mathrm{dif}^2c$
in both the acceleration and diffusion zones, 
with $L_s(\epsilon)$ denoting the synchrotron Luminosity as a function of photon energy in units
of electron rest energy.
The same energy density caused by the acceleration zone is more inhomogeneous, and approximated as
$3L_\mathrm{s,acc}(\epsilon)/4\pi R_\mathrm{acc}^2c$ in the acceleration zone, 
and $3L_\mathrm{s,acc}(\epsilon)/4\pi R_\mathrm{dif}^2c$ in the diffusion zone.
\footnote{Since the considered acceleration zone is much smaller than the diffusion zone, 
we use $R_\mathrm{dif}\simeq R_\mathrm{em}$ and $t_\mathrm{esc,dif}\simeq t_\mathrm{esc,em}$.}
To match the cases we study in the 2D
model, we choose $R_\mathrm{dif}=8.25R_\mathrm{acc}$.

The calculation of the steady-state electron spectrum is described in Appendix A.
Four time scales, namely the acceleration time scale $t_\mathrm{acc}$, the cooling time scale $t_\mathrm{cool}$,
the escape time scale from the acceleration region ,$t_\mathrm{esc,acc}$, and the emission region, $t_\mathrm{esc,em}$,
are important in determining the total electron energy distribution (EED). 
As illustrated in Fig.\,\ref{fig:tscale}, the Lorentz factor $\gamma_\mathrm{max}$,
at which the high-energy cut off starts, is determined by a balance between $t_\mathrm{acc}$ and $t_\mathrm{cool}$; the
Lorentz factor $\gamma_\mathrm{b}$, at which there is a spectral break, is determined by the relationship between
$t_\mathrm{esc,em}$ and $t_\mathrm{cool}$; the spectral index of the EED above the spectral break, p, is determined
by the ratio between $t_\mathrm{acc}$ and $t_\mathrm{esc,acc}$.

The synchrotron power and synchrotron spectrum are calculated in the same way as we will do in the 
2D model \citep{chen_etal:2011:multizone_code_mrk421}.
We follow 
\citet{graff_etal:2008:pipe} of using the $\delta$-function appoximation to get the IC
emission through a simple integration, 
\ie $\epsilon_\textsc{ic}=\frac{4}{3}\gamma^2\epsilon_0$,
and using a step-function approximation for the Klein-Nishina effect 
(Thomson scattering for $\gamma \epsilon_0 < 3/4$; no scattering for $\gamma \epsilon_0 \geq 3/4$ ).
With this approach, we integrate over the seed photon distribution to obtain
\begin{equation}
\begin{split}
& j_\textsc{ic}(\epsilon_\textsc{ic},t) = (\frac{\sqrt{3}}{4}\sigma_\textsc{t} c \sqrt{\epsilon_\textsc{ic}}) \\
& \times \int_0^{\epsilon_\text{max}} n(\gamma,t) U(\epsilon_0,t)
\epsilon_0^{-3/2}d\epsilon_0, \ \ \gamma=\sqrt{\frac{3\epsilon_\textsc{ic}}{4\epsilon_0}}
\end{split}
\end{equation}
where
\begin{equation}
\epsilon_\text{max} =
\begin{dcases}
 \frac{3\epsilon_\textsc{ic}}{4},  \text{when} \ \epsilon_\textsc{ic}\leq1, \\
 \frac{3}{4\epsilon_\textsc{ic}},  \text{when} \ \epsilon_\textsc{ic}>1,
\end{dcases} 
\end{equation}
because of the Klein-Nishina effect.

The resulting EEDs have a sharp cutoff at the highest energy. This is caused by the 
simplification of not considering radiative cooling in computing the electron spectrum in the acceleration 
zone.
A direct high-energy cutoff on the particle spectrum in the acceleration zone
is implemented based on a posterior consideration of the cooling.
Since this cutoff also affects the particle number, especially when the spectrum is hard, we make
a correction to the particle number density afterwards.
This ensures that with the particle escape and particle injection considered, 
the total particle number is conserved ($n_\mathrm{e,acc}/t_\mathrm{esc,acc}=Q$).

\begin{table}
\centering
\caption{The parameters used for the benchmark cases. The observation angle is always $1/\Gamma$ so that
the Doppler factor $\delta$ is equal to the bulk Lorentz factor $\Gamma$.
The volume hight $Z=4R/3$ in all 2D cases. $D_\mathrm{x}$, $t_\mathrm{esc,acc}$ and $t_\mathrm{esc,dif}$ are not independent
in the 2D model. Here $D_\mathrm{x}$ is an input parameter, while the other two are measured when the
simulation reaches the steady state.}
\label{tab:par}       
\begin{tabular}{llll}
\hline
 & two zone & closed & open \\\hline
B(G) & 0.3 & 0.3 & 0.3 \\
$\delta$ & 33 & 33 & 33 \\
$\gamma_\mathrm{inj}$ & 33 & 33 & 33 \\
$R(cm)$ & $0.75\times10^{16}$ & $0.75\times10^{16}$ & $0.75\times10^{16}$ \\
$n_\mathrm{e}(cm^{-3})$ & 2.07 & 30 & 2.07 \\
$t_\mathrm{acc}(R/c)$ & 0.267 & 0.4 & 0.267 \\
$D_\mathrm{x}(cm^2/s)$  & - & $1.9\times10^{24}$ & $3.8\times10^{24}$ \\
$t_\mathrm{esc,acc}(10^5s)$ & 0.84 & - & 0.84 \\
$t_\mathrm{esc,dif}(10^5s)$ & 23.3 & - & 23.3 \\
\hline
\end{tabular}
\end{table}

We guide our modeling using the SEDs of Mrk 421. But we restrict ourselves from matching the SEDs in details, 
to avoid excessive time spent on fine tuning of parameters. 
We also intend to keep our results generally applicable to different objects.

\begin{figure*}
\includegraphics[width=0.49\linewidth]{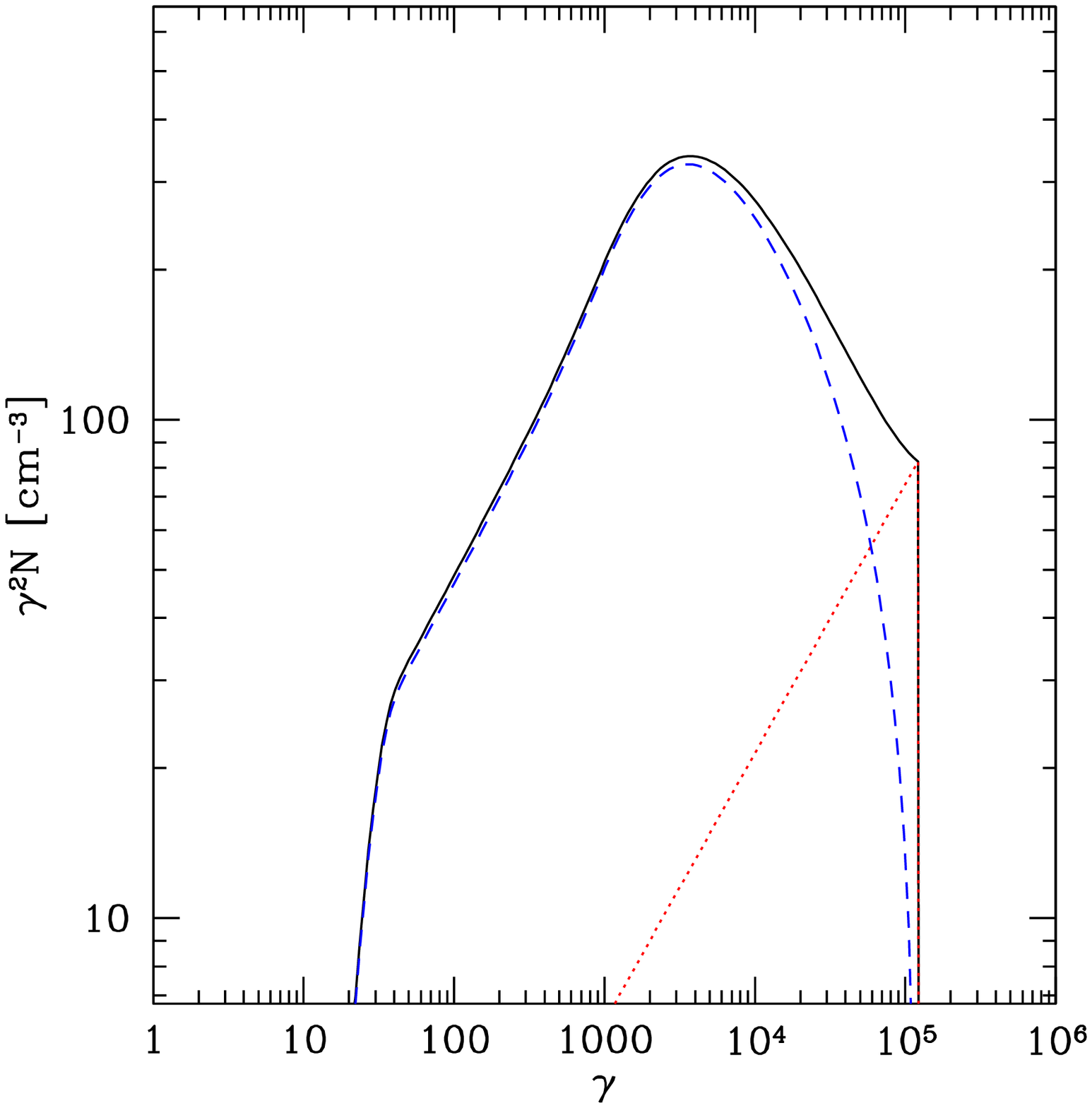}
\includegraphics[width=0.49\linewidth]{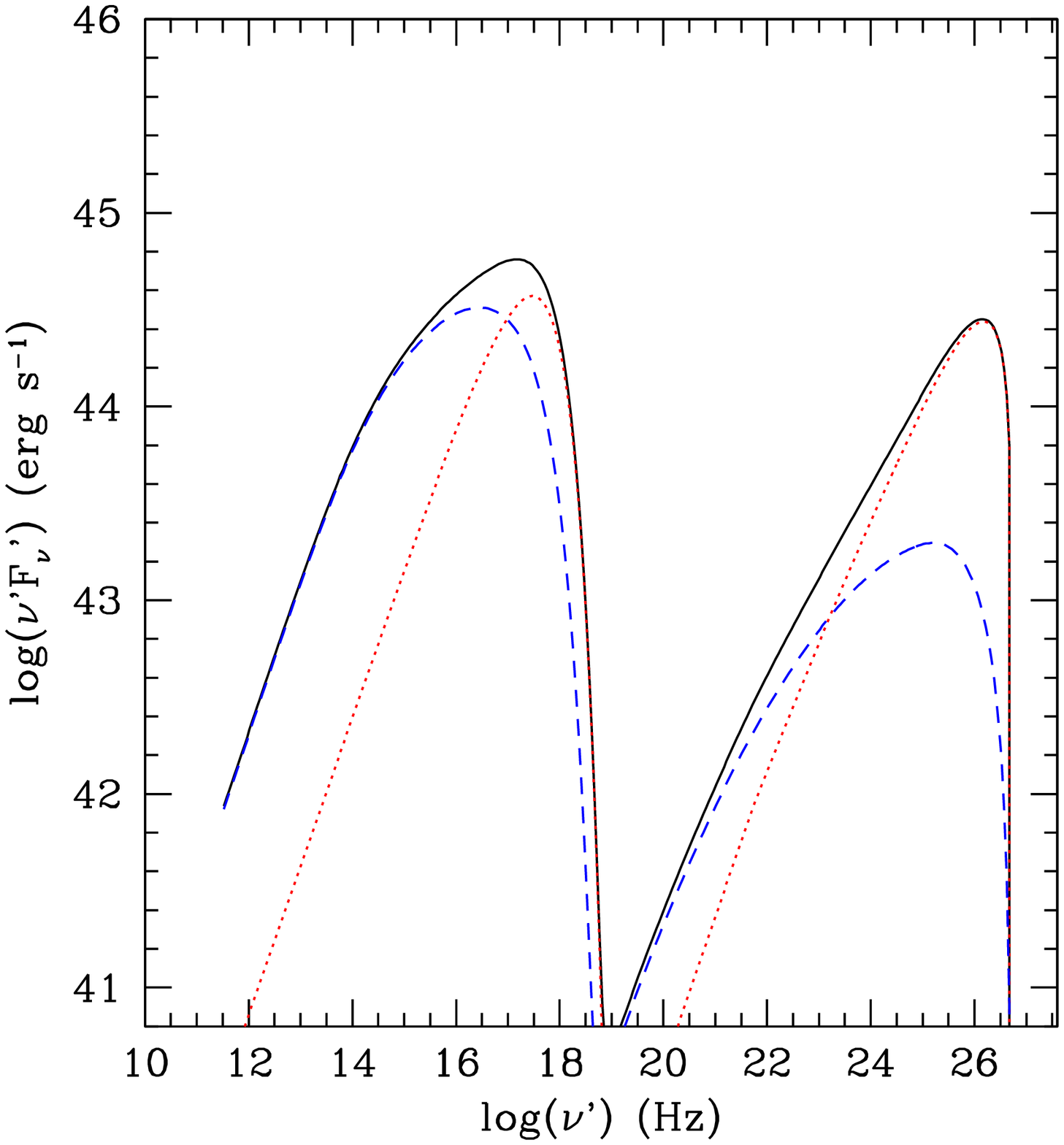}
\caption{The EED and SED from the two-zone model. The parameters are chosen to match those of 
one case of the 2D model that will be discussed in \S \ref{esc:center}.
The total EED and SED are plotted with black solid lines, while the contributions from
the acceleration zone and the diffusion zone are plotted in red dotted lines and blue dashed lines respectively.
The spectral indices of the EED are -1.42 at $\gamma=2\times10^2$ and -2.52 at $\gamma=2\times10^4$.
The spectral indices of the SED are -0.68 at 10\,eV ($2.42\times10^{15}Hz$) and -0.52 at 1\,GeV ($2.42\times10^{23}Hz$).
}
\label{fig:2zonect}
\end{figure*}

In a benchmark case for the two-zone model (Fig.\,\ref{fig:2zonect}, parameters listed in Table \ref{tab:par}), the EED forms a typical broken power-law distribution, 
with a spectral break of $\sim1$ at $\gamma\approx3\times10^3$. Because there is a concentration of higher-energy
synchrotron photons in the acceleration zone, emitted by the higher-energy electrons in that same zone,
the seed-photon field for the SSC is disproportionately strong for the highest-energy electrons.
This preference of SSC scattering between the high-energy electrons and the high-energy photons causes the SSC
spectrum to be harder than the synchrotron spectrum, especially at frequency below the SED peaks, 
above which Klein-Nishina effect begins to play a role.
This effect is clearly visible in Fig.\,\ref{fig:2zonect} right, where the spectral indices are measured to be
-0.68 at 10\,eV and -0.52 at 1\,GeV.

\begin{figure*}
\includegraphics[width=0.49\linewidth]{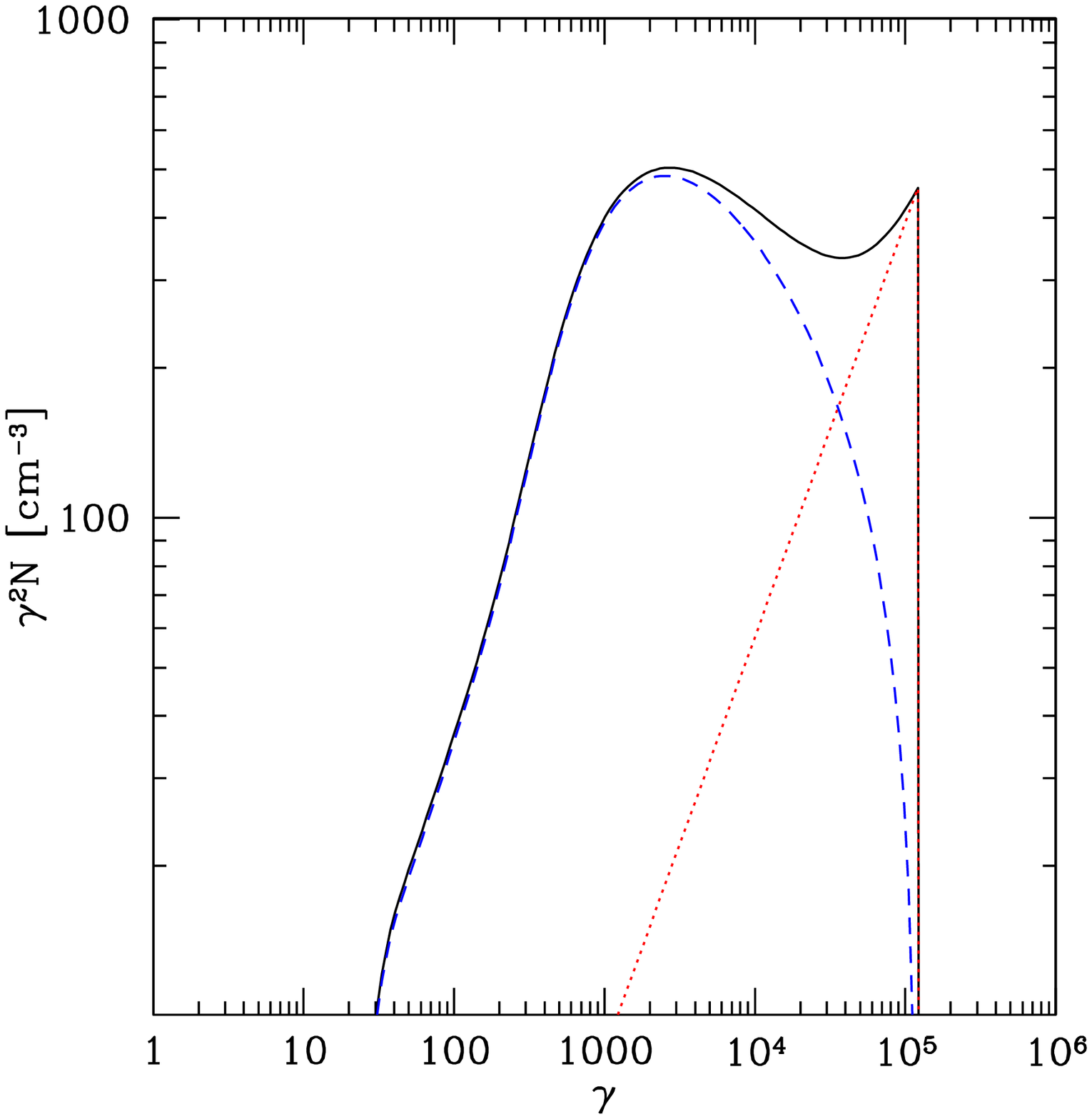}
\includegraphics[width=0.49\linewidth]{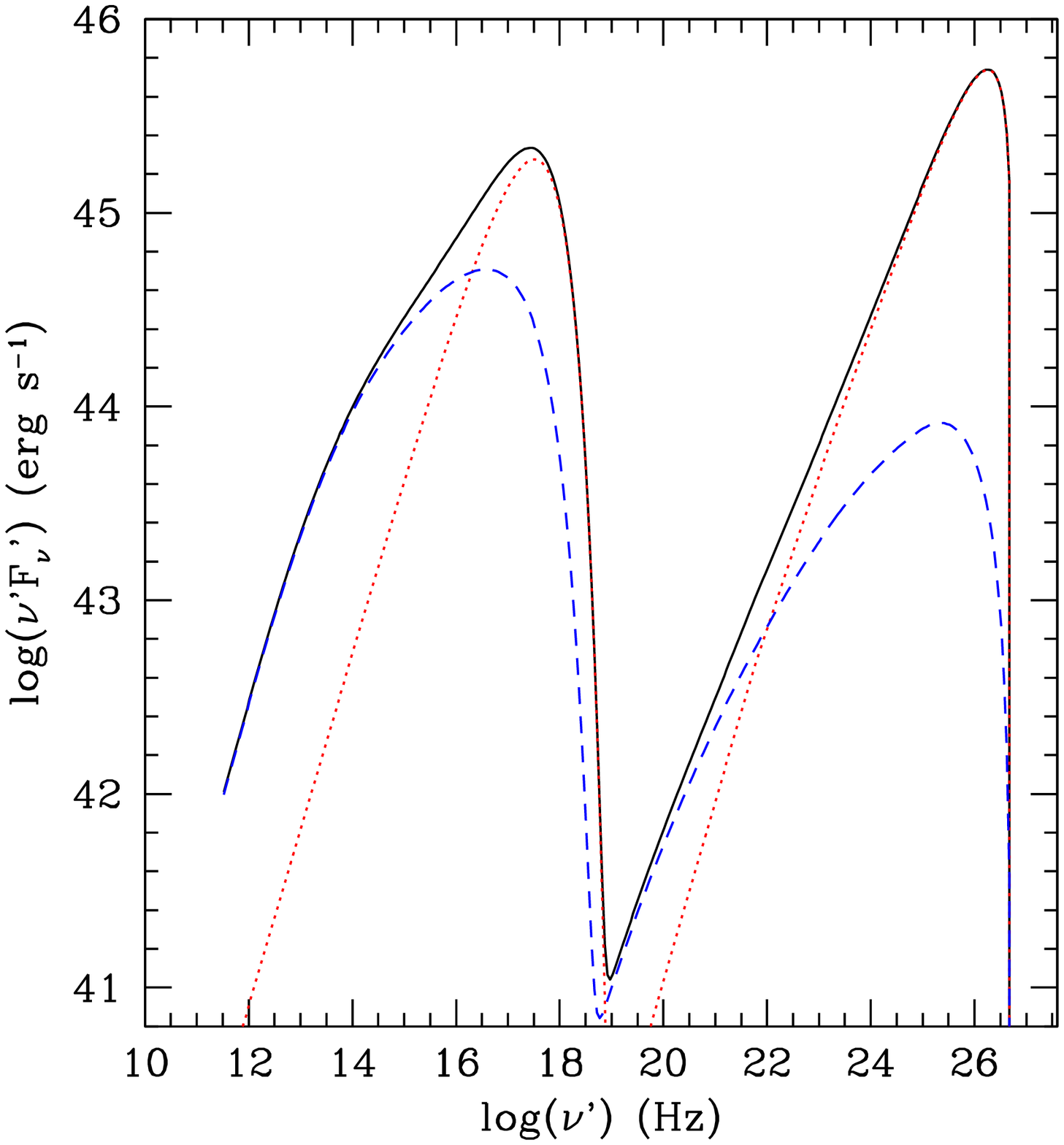}
\caption{The EED and SED from the two-zone model, with relatively slow particle escape. 
The parameters are chosen to match those of 
one case of the 2D model that will be discussed in \S \ref{esc:slow}. The color scheme is the same as Fig.\,\ref{fig:2zonect}.
The spectral indices of the SED are -0.597 at 10\,eV ($2.42\times10^{15}$Hz), -0.587 at 50\,eV ($1.21\times10^{16}$Hz) and -0.326 at 10\,GeV ($2.42\times10^{15}$Hz).}
\label{fig:2zonebump}
\end{figure*}

In another case (Fig.\,\ref{fig:2zonebump}) the particle escape time is three times longer.
This results in a harder electron spectrum, which leads to a dominance in the spectrum at the highest energy 
by electrons in the acceleration zone.
Looking at the EED from low energy to high energy, this shift of dominance causes a spectral hardening
at the highest energy, because the un-cooled electron spectrum in the acceleration zone is harder than
the cooled electron spectrum in the diffusion zone.
This feature is clearly visible in Fig.\,\ref{fig:2zonebump} left. But it is less apparent in 
Fig.\,\ref{fig:2zonebump} right, because the SED is similar to a $\gamma^3N-\gamma$ representation, 
instead of the EED shown on the left which is a $\gamma^2N-\gamma$ representation.
A careful examination of the synchrotron spectral index reveals that a slight hardening of the spectrum
by 0.01 is still present in the synchrotron SED.
A combination of this EED hardening and the above mentioned hardening of the SSC spectrum result in a
very hard GeV spectrum with spectral index of about -0.3 (equivalent to a photon index of -1.3).

\section{The 2D model}
\label{model}

The semi-analytic two-zone model already shows some unique spectra features that are not captured in 
one-zone models.
However, there are some significant simplifications in the analytic approach that limits
the accuracy of the model, \eg the neglect of radiative cooling in the acceleration zone, 
the return-flux for lower-energy particles from the diffusion zone to the acceleration zone, 
and the inhomogeneity within the acceleration and diffusion zones.
Further more, the applicability of the analytic model is limited because it does not account for the
LTTE, which is important especially in studies of variability.

Taking one step beyond the two-zone analytic model,
in this section we will describe our time-dependent 2D numerical model we built
to study the particle acceleration and spatial diffusion in inhomogeneous jets.
We consider a 2-dimensional axisymmetric jet model that is built on the
Monte-Carlo/Fokker-Planck (MCFP) code developed by \citet{chen_etal:2011:multizone_code_mrk421}.
This model employs an approach combining the Monte-Carlo (MC) method for photon tracking and
scattering, and Fokker-Planck (FP) equation for the electron momentum evolution (hence the name MCFP).
The full transport equation takes the form
\begin{equation}
\label{eq:FPeq}
\begin{split}
 \frac{\partial n (\gamma,\mathbf{r},t)}{\partial t} & =
-\frac{\partial}{\partial \gamma}\bigg[n(\gamma,\mathbf{r},t)\dot{\gamma}(\gamma,\mathbf{r},t)\bigg] \\
 &  +\frac{\partial}{\partial \gamma}\bigg[D(\gamma,\mathbf{r},t)
 \frac{\partial n(\gamma,\mathbf{r},t)}{\partial \gamma} \bigg] + Q(\gamma,\mathbf{r},t) \\
& -\mathbf{\nabla}\cdot\bigg[D_x(\gamma) \nabla n(\gamma,\mathbf{r},t)\bigg] .
\end{split}
\end{equation}
$n(\gamma,\mathbf{r},t)$ is the differential number density of particles. 
The first term on the right hand side
\begin{equation}
\label{eq:acc}
\dot{\gamma}(\gamma,\mathbf{r},t)=\dot{\gamma}_\mathrm{cool}(\gamma,\mathbf{r},t)
+\dot{\gamma}_{D}(\gamma,\mathbf{r},t) \ ,
\end{equation}
includes both radiative cooling $\dot{\gamma}_\mathrm{cool}(\gamma,\mathbf{r},t)$ 
and stochastic acceleration  
$\dot{\gamma}_{D}(\gamma,\mathbf{r},t)=\frac{\gamma}{t_\mathrm{acc}}$ in the acceleration region, caused by momentum diffusion of particles.
The dispersion effect of the diffusion is described by the second term, also applicable in the acceleration region only, where the diffusion coefficient is
\begin{equation}
D(\gamma,\mathbf{r},t)=\frac{\gamma^2}{2 t_\mathrm{acc}}.
\end{equation}
The third term represents the injection of particles.
The fourth term is the spatial diffusion of particles.
$D_x(\gamma)$ is the spatial diffusion coefficient. $D_x(\gamma)$ could easily
be energy dependent in our calculation, but in this work we restrict our discussion to the energy independent situations
to reduce the number of free parameters.
This also implies that the momentum diffusion coefficient, 
which is associated with the spatial diffusion, should be proportional to $\gamma^2$, 
\ie, $t_\mathrm{acc}$ should be energy independent \citep{shalchi_2012:turbulence_stochastic_acc.19.10}.
Also, only under this assumption is the analytical solution used in the two-zone model available 
(See Appendix A). Restricting the 2D model to this assumption makes the comparison of the two models 
much easier.
More discussion on energy dependent $t_\mathrm{acc}$ can be found in \citet{tramacere_etal:stochastic_acceleration:2011}.

We use operator splitting to treat the momentum terms and spatial terms separately.
Without the spatial terms, the equation is reduced to the FP equation. The finite-difference method used
to solve the FP equation is described in details in \citet{chen_etal:2011:multizone_code_mrk421}.
The spatial terms of the transport equation is handled using the finite-element method, where we calculate
the flux at each spatial boundary, and use those fluxes to update the density in each cell.

The diffusion causes propagation of particles to neighboring cells at every time step.
The time step is set to be the light-crossing time of a single cell $z/c$, so the speed of particle
escape and information exchange does not exceed the speed of light.

With spatial diffusion considered, we focus on the effects of localized particle acceleration,
\ie, the acceleration region is a small accelerator.
This acceleration region can occupy either one or multiple cells in the 2D model, but is conceptually
equivalent to the acceleration zone in the two-zone model.
This acceleration can represent either second-order Fermi acceleration or acceleration by magnetic reconnection,in both of which the acceleration could be restricted to small turbulence regions.

\begin{figure}
\centering
\includegraphics[width=0.99\linewidth]{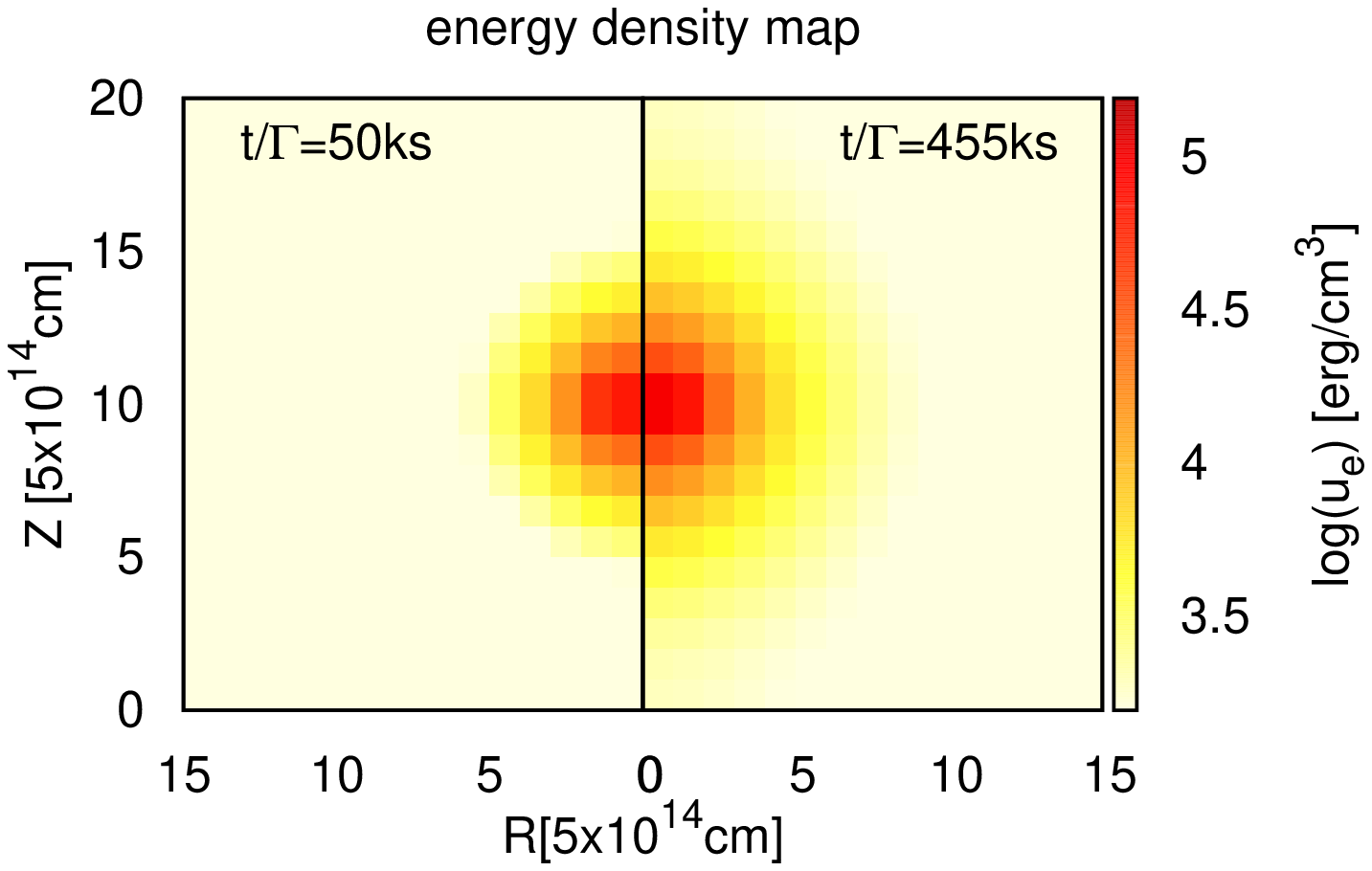}
\includegraphics[width=0.99\linewidth]{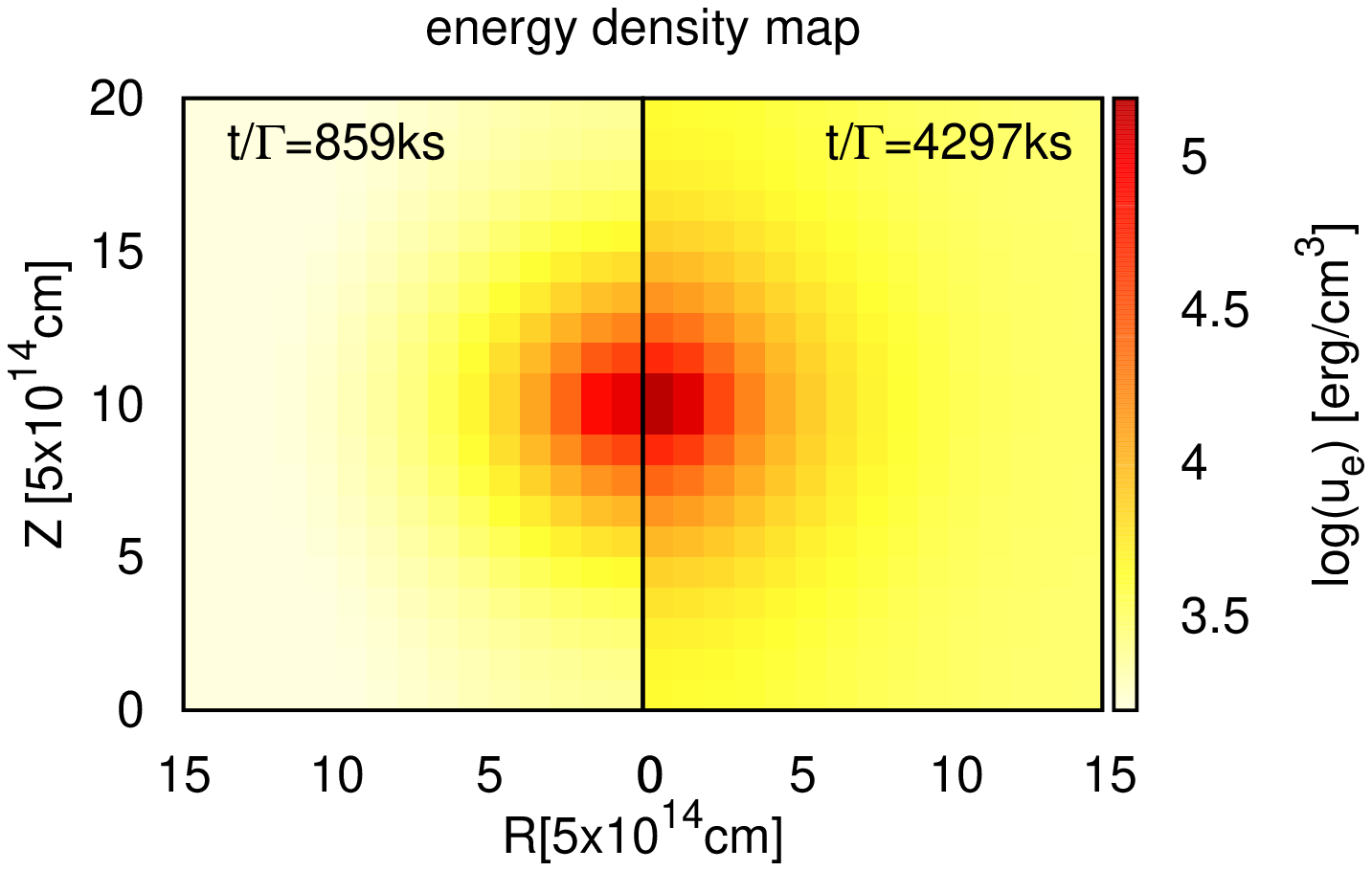}
\caption{Energy density maps of the confined particle diffusion case at simulation time step 100, 900, 1700 and 8500
(corresponding to the electron energy distribution in Fig.\,\ref{fig:refl_ct_maps} with color green, red, black and cyan).
The color scheme spans two order of magnitude, and is normalized
so that the highest density in the last figure is represented by dark red. 
The left/right halves of the figures show the maps at different time.
The spatial unit ($5\times10^{14}$cm)
is chosen to be equal to the grid size of the simulation.
}
\label{fig:refl_ct_4emaps}       
\end{figure}

In the 2D model we consider scenarios with reflecting (closed) boundary condition 
(zero flux between the surface cells $r=r_{max}, z=1, z=z_{max}$ and the imaginary cells outside of 
the emission region) 
and escape (open) boundary condition 
($N(\gamma,\mathbf{r},t)=0$ at the imaginary cells; reflecting boundary condition is always used for the innermost boundary). 
In the former case the particle number is conserved,
so the system will reach a steady state after a while.
In the later case, the particles keep escaping from the emission region,
with an implicit assumption that any particle outside of the emission region has negligible contribution to the emission.
This is true for synchrotron radiation, if the magnetic field in outer regions are much weaker.
It is also valid for SSC emission because of the lower synchrotron radiation density resulted from the weaker magnetic field.
However, for EC emission, this assumption needs more careful examination.
With particles continuously escape from the emission region,
steady state only exists when there is additional source of particle pick up. This may happen at
the same locations as the particle acceleration, because the turbulent magnetic field there may trap 
and isotropize particles in the intergalactic medium.
In those cases, the trapped particles may have Lorentz factor similar to the bulk Lorentz factor of
the jet. This is how we choose $\gamma_\mathrm{inj}$ in our models. Effectively,this particle Lorentz factor
determines the minimum Lorentz factor of the steady state EED in the open boundary scenario.
In both the closed and open boundary scenarios, initially the emission regions have homogeneous 
particle distributions that form a power-law EEDs with spectral index -1.1 between Lorentz factor 1 and 33.
This choice of initial particle distribution only affects the early evolution of the EED, 
it hardly has any effect on the final steady state spectra.

In this paper, our discussion focuses on SSC scenarios, even though some results may be generalized to EC scenarios as well, especially the results with closed boundary conditions, 
or if synchrotron emission is the primary subject of concern.
Synchrotron-self absorption is also included in the 2D model, 
although it turns out to be not very important above 2 GHz in the cases discussed in this paper.
All the simulation shown in this work use 20 layers in the longitudinal direction and 15 layers in the radial direction (nz=20 and nr=15). The length/radius ratio (Z/R) is 4/3, so the cell
sizes in longitudal and radial direction (dz and dr) are the same. The simulation time step is chosen to be
the same as the light crossing time of one cell (dz/c).

\begin{figure*}
\centering
\includegraphics[width=0.33\linewidth]{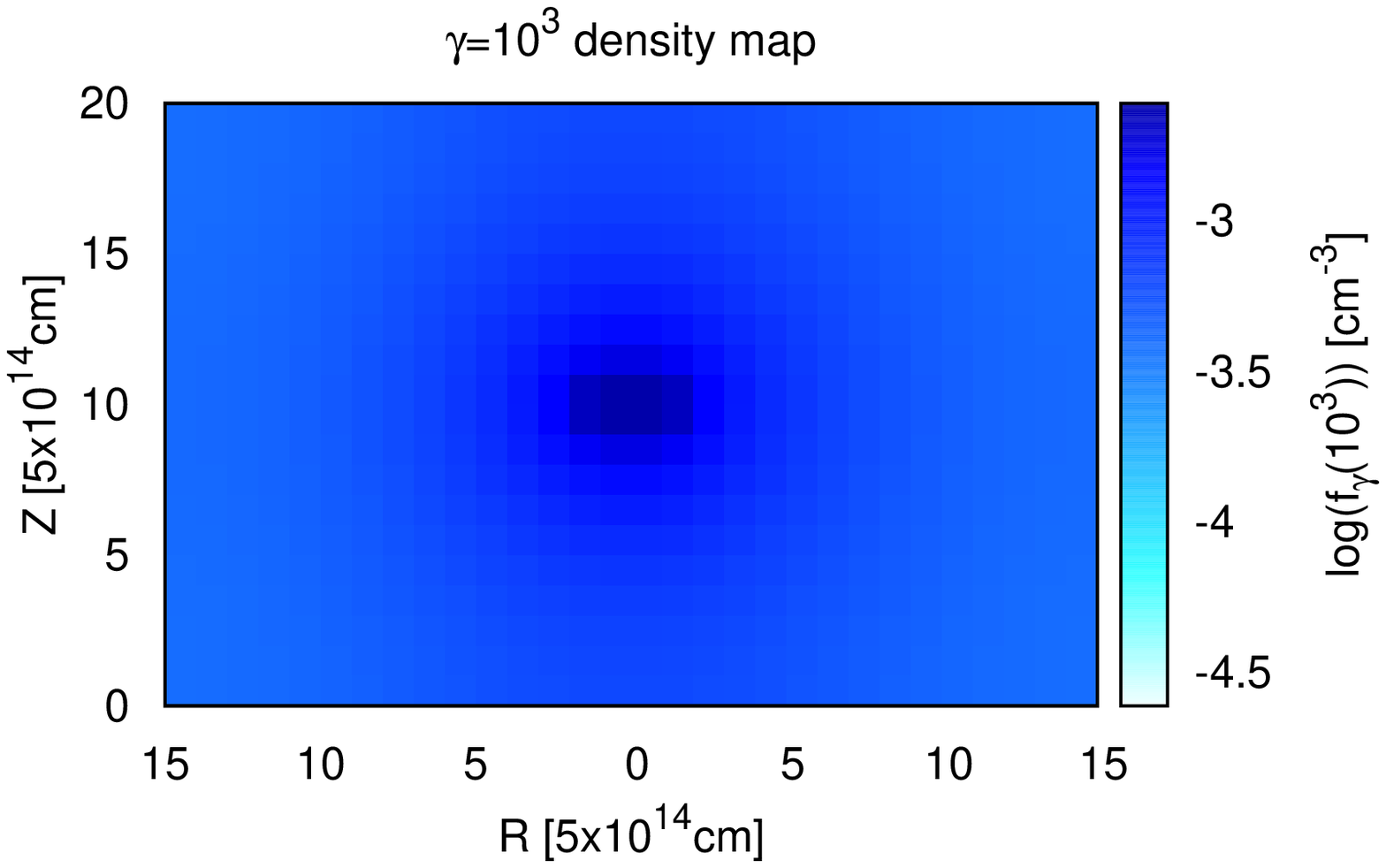}
\includegraphics[width=0.33\linewidth]{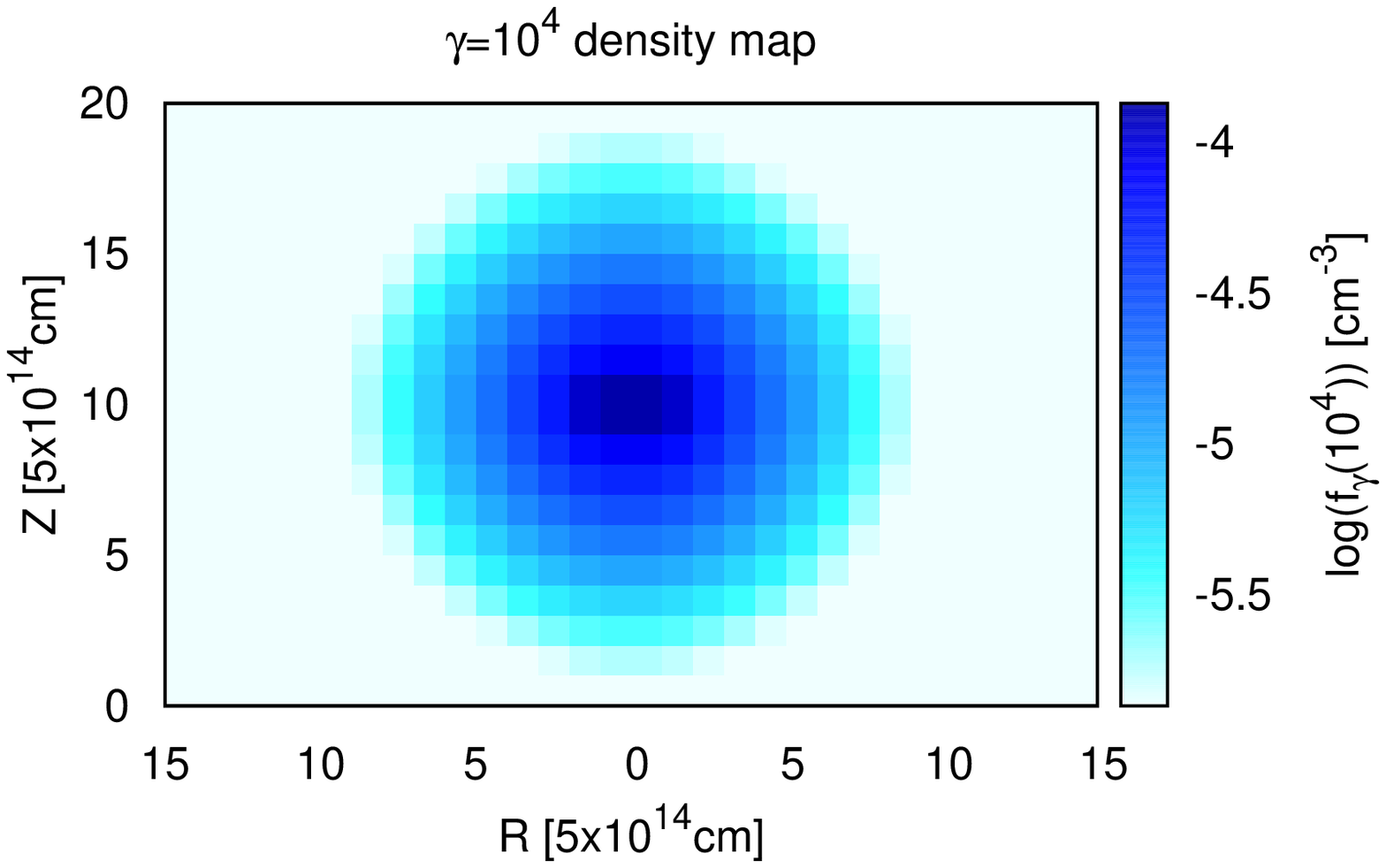}
\includegraphics[width=0.33\linewidth]{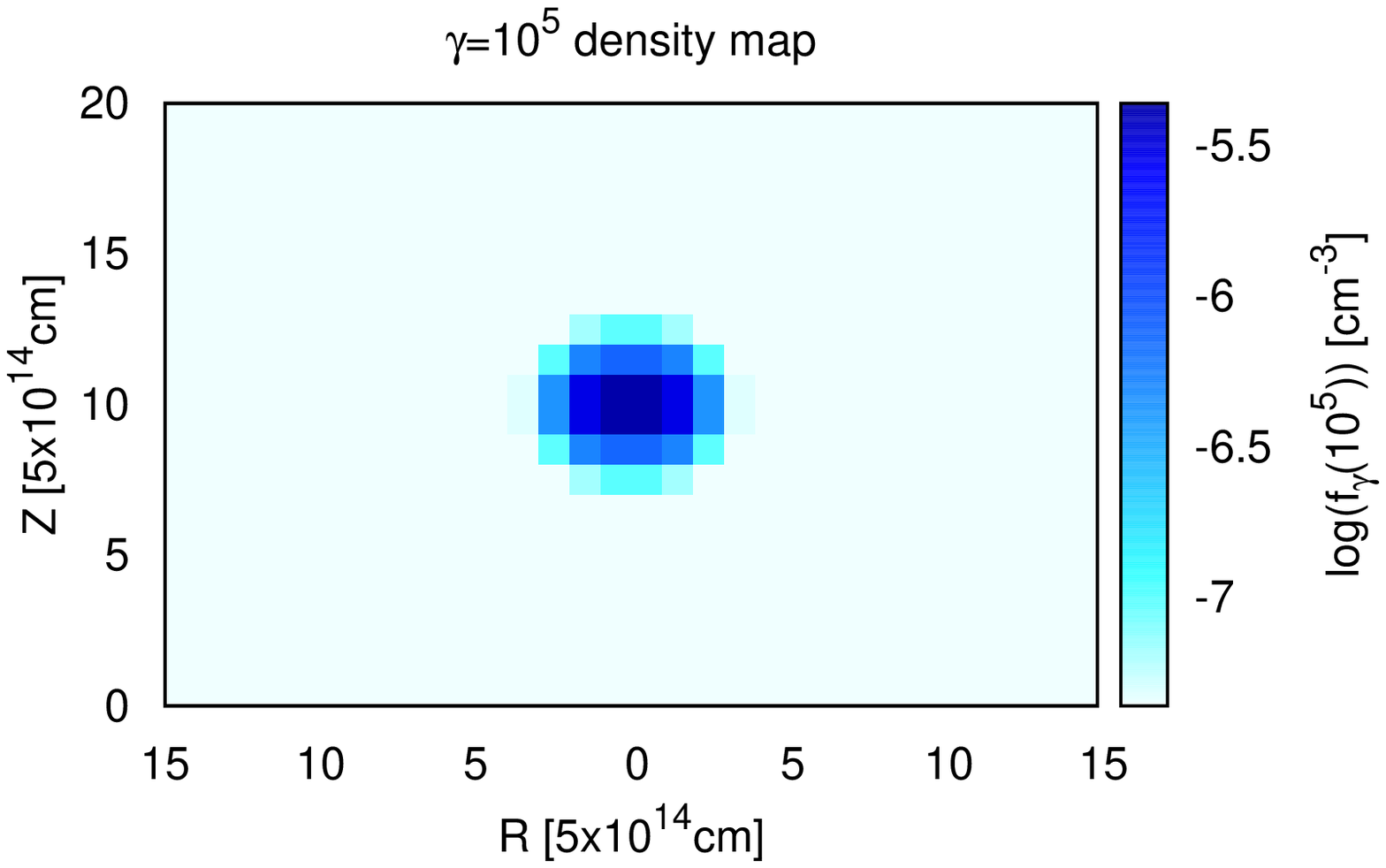}
\caption{Density maps of the confined particle diffusion case at simulation time step 1700.
They are  maps of differential density $N_\gamma$ for electrons
with Lorentz factor $10^3$, $10^4$ and $10^5$. 
The left halves of the images are mirror image
of the right halves to illustrate the cylindrical geometry. }
\label{fig:refl_ct_maps}       
\end{figure*}

\begin{figure*}
\centering
\includegraphics[width=0.49\linewidth]{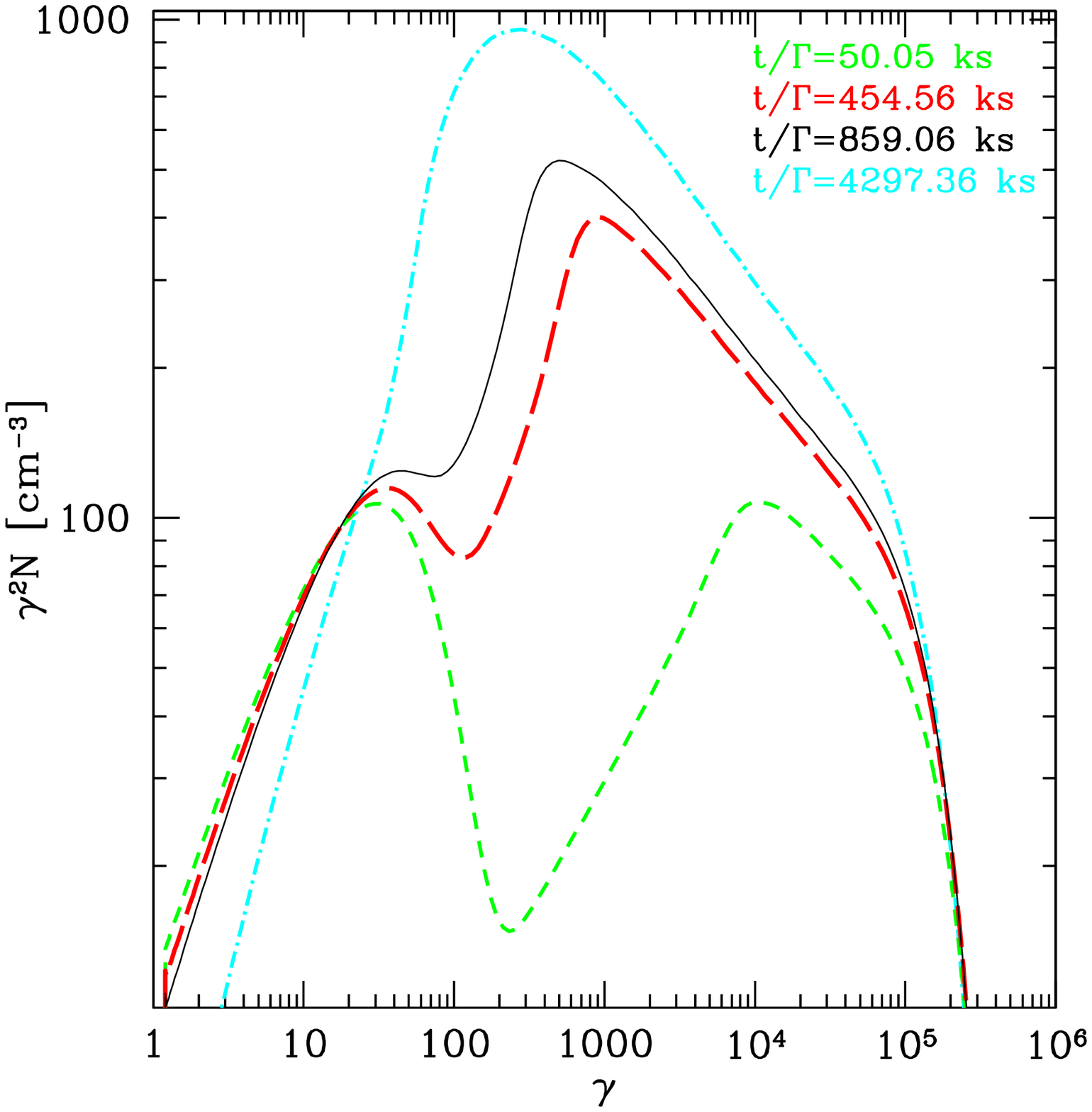}
\includegraphics[width=0.49\linewidth]{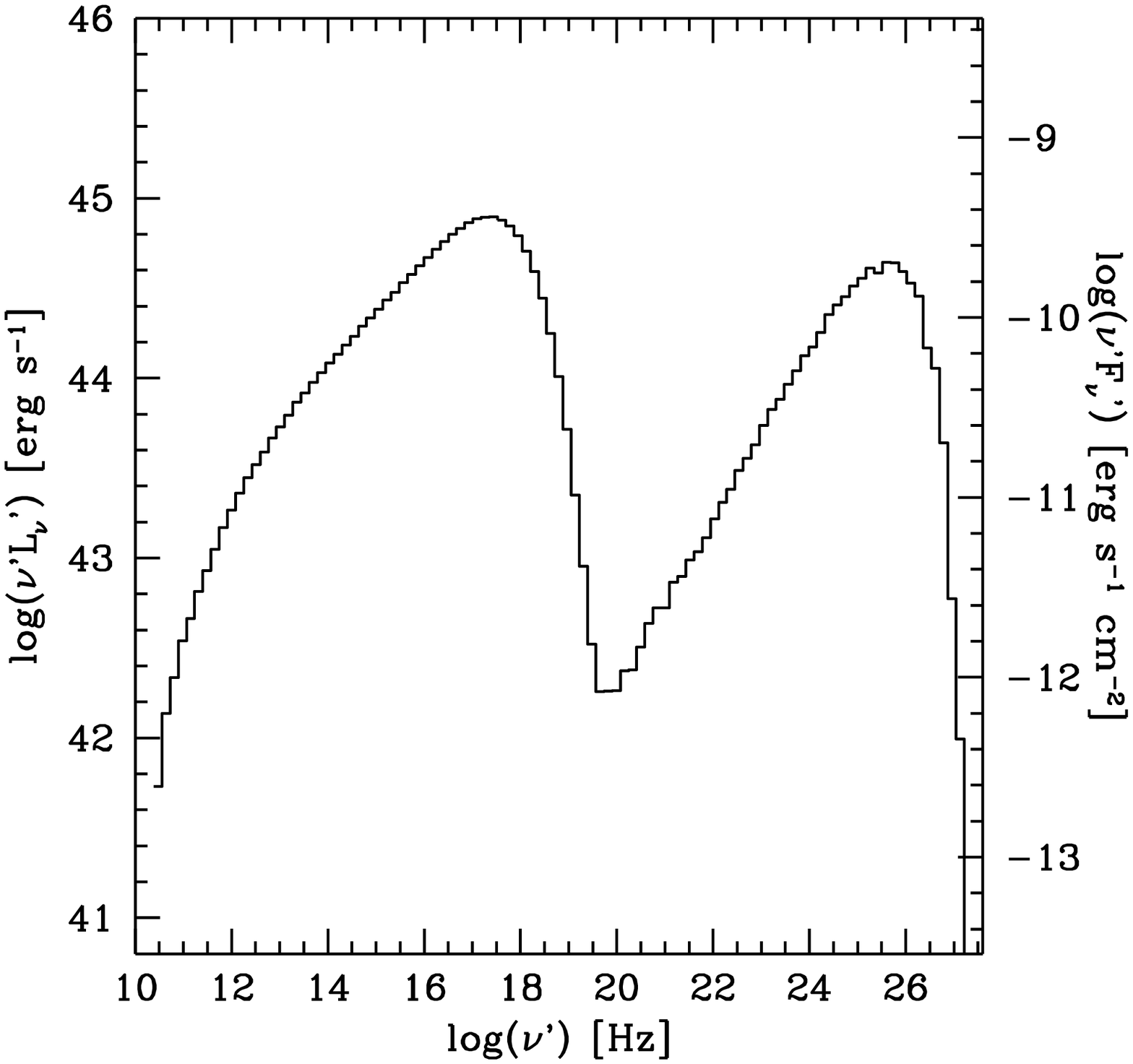}
\includegraphics[width=0.33\linewidth]{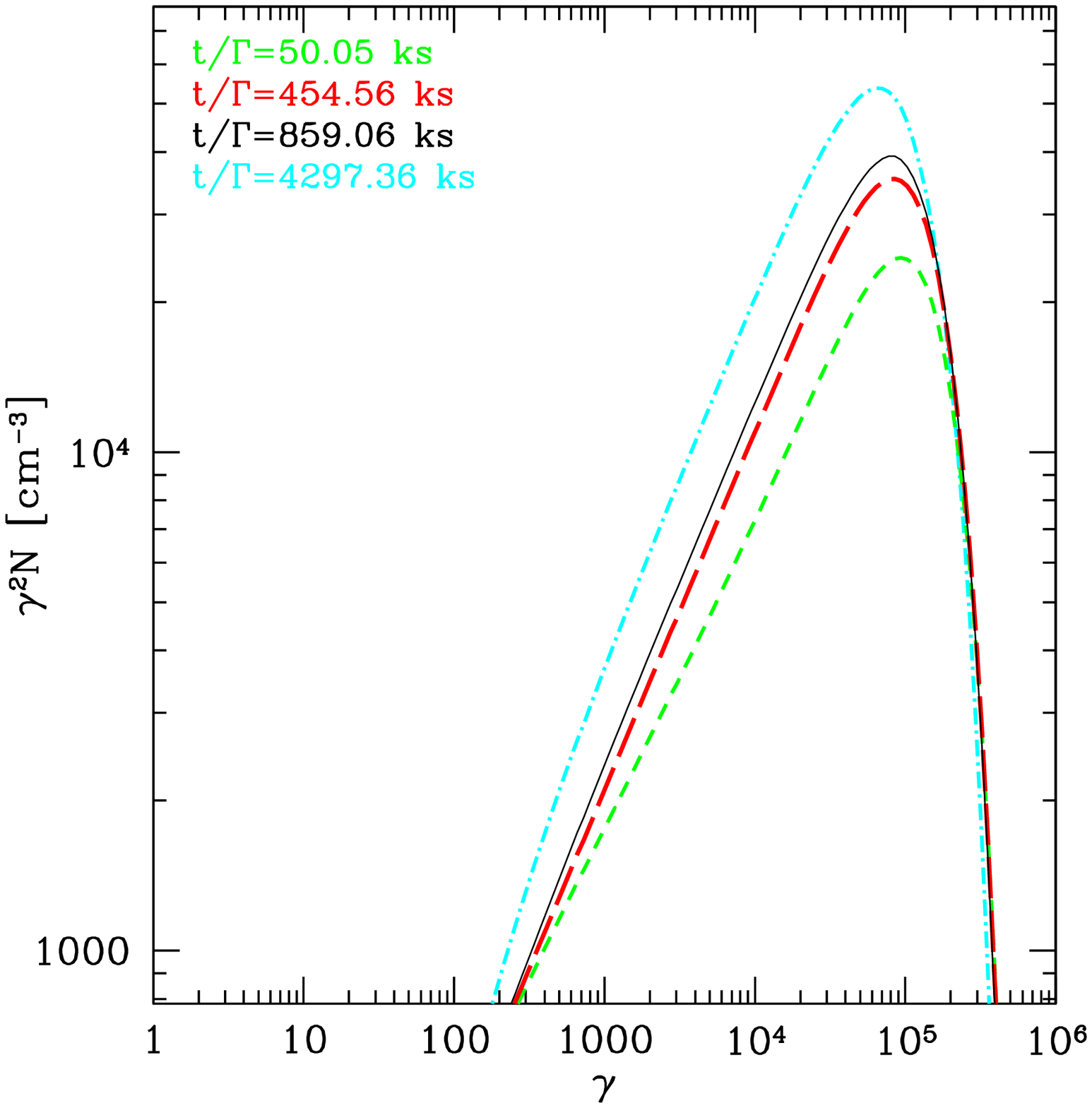}
\includegraphics[width=0.33\linewidth]{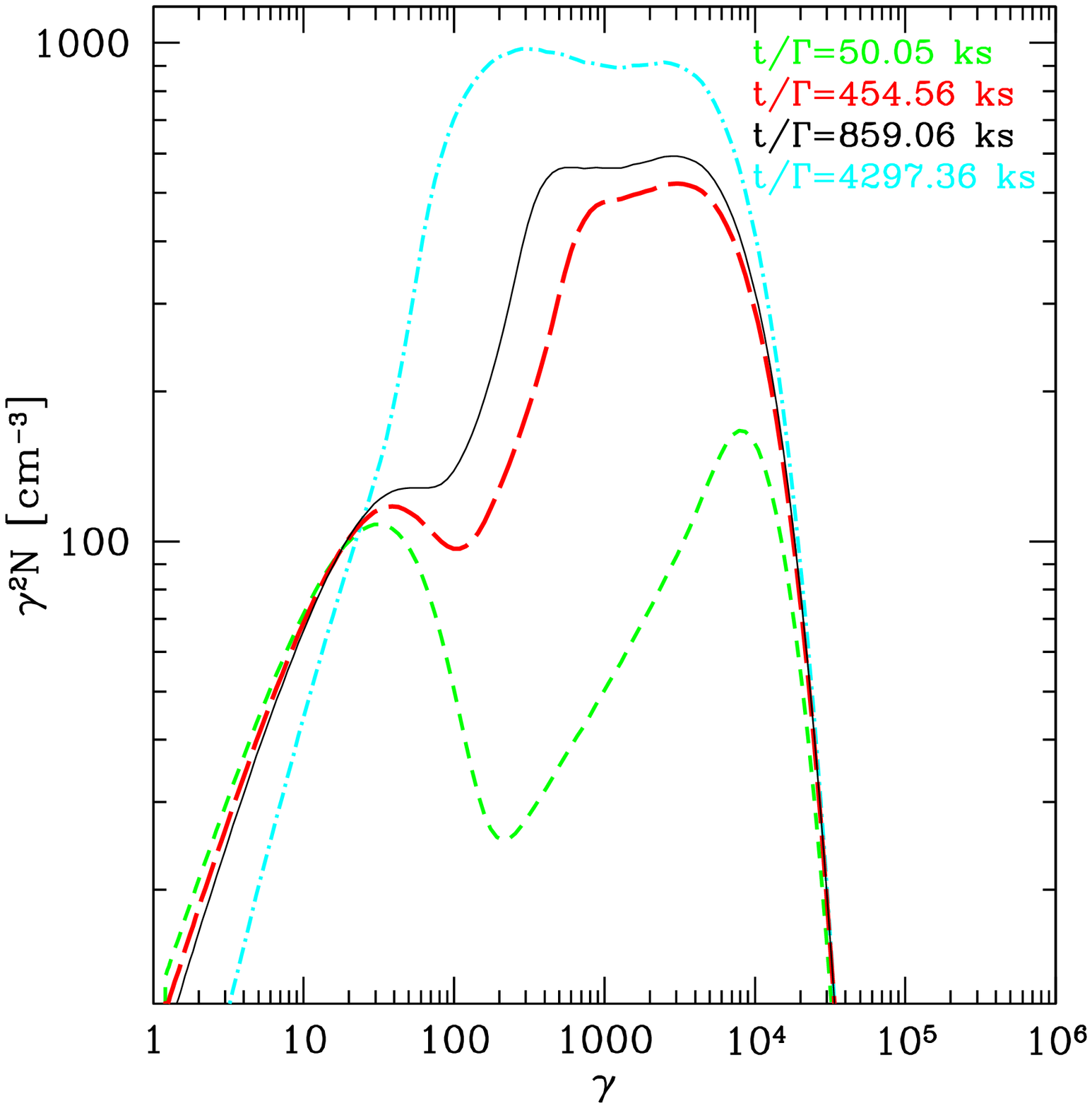}
\includegraphics[width=0.33\linewidth]{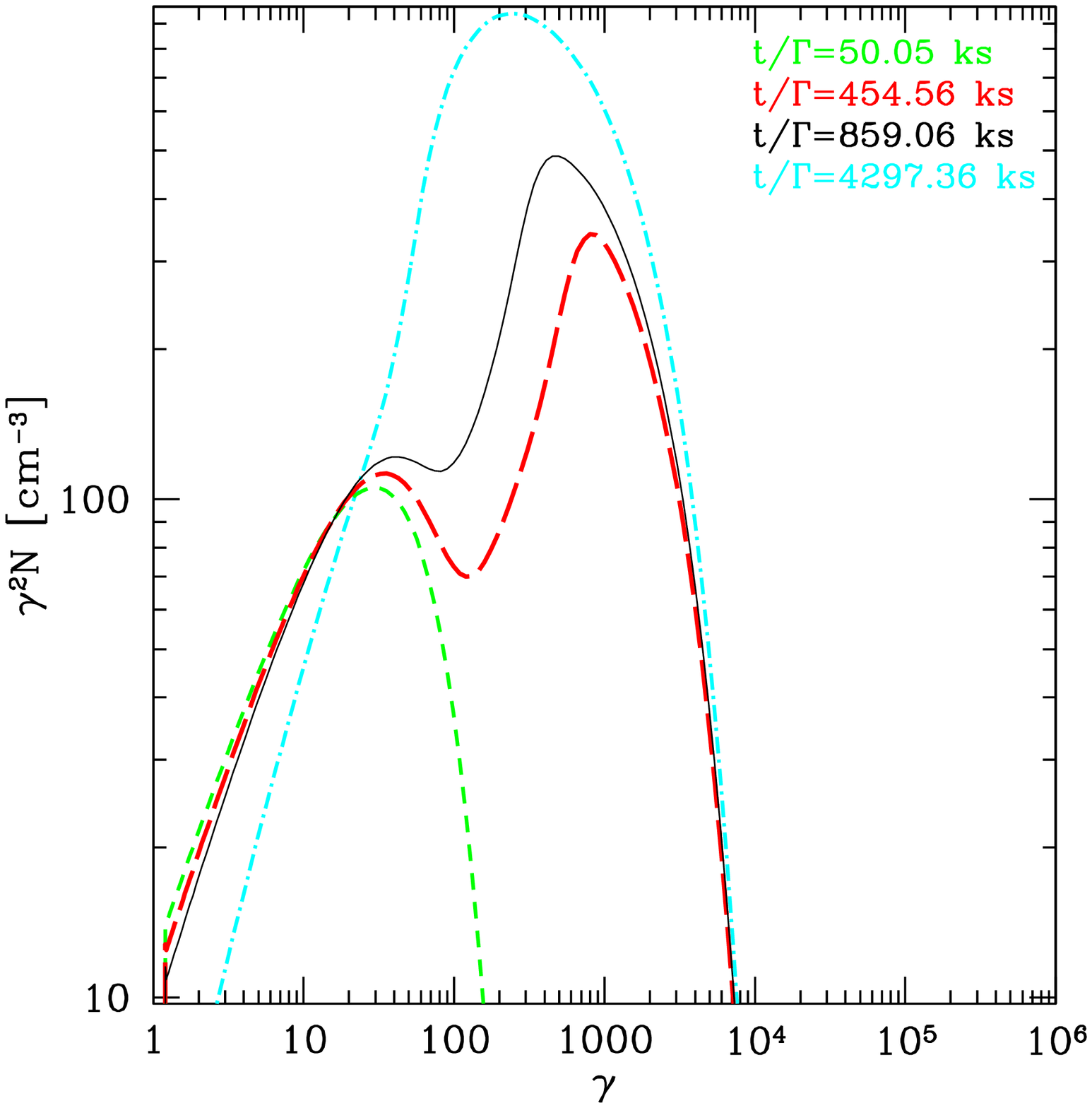}
\caption{(Top) Total electron energy distribution (EED) and spectral energy distribution (SED).
The SED have spectral indices -0.74 at 10\,eV, and -0.67 at 1\,GeV.
(Bottom) The EED by individual cells. All of them are from the middle
secion in the longitudal direction (jz=11). Radially, they are (from left to right) from the inner ,middle and outter cells (kr=1,8,15).
}
\label{fig:refl_ct_eed}
\end{figure*}

\section{2D Results}
\label{results}
With this time-dependent 2D model, we study the acceleration and diffusion of particles,
as well as their synchrotron and IC emissions. The different cases we study, along with the 
associated section and figure numbers, are listed in Table \ref{tab:case}.

\begin{table}
\centering
\caption{}
\label{tab:case}       
\begin{tabular}{lll}
\hline
boundary: & closed & open \\\hline
centered acc. & \S \ref{refl:center}, Fig.\,\ref{fig:refl_ct_4emaps}, \ref{fig:refl_ct_maps} \& \ref{fig:refl_ct_eed} 
& \S \ref{esc:center}, Fig.\,\ref{fig:esc_ct_maps} \& \ref{fig:esc_ct_eed}  \\

slow diffusion & \S \ref{refl:slow}, Fig.\,\ref{fig:refl_bump} 
& \S \ref{esc:slow}, Fig.\,\ref{fig:esc_bump}  \\

off-center acc. & - 
& \S \ref{esc:away}, Fig.\,\ref{fig:esc_away}, \,\ref{fig:esc_away2} \& \ref{fig:esc_ctaway} \\

elongated acc. & -  
& \S \ref{esc:long}, Fig.\,\ref{fig:esc_long}  \\

\hline
\end{tabular}
\end{table}

\subsection{Confined Particle Diffusion}
\label{reflection}
In this section we discuss a closed emission volume, in which the particle diffuse spatially within
a confined cylindrical region, while there is no particle exchange/escape at the outer-most boundary,
\ie the total particle number is conserved. 

Similar cases discussed in this section cannot be studied with our two-zone model in \S \ref{2zone},
because in the two-zone mode we did not include the backflow of low energy particles from the diffusion region
to the acceleration region. This backflow is important in the closed boundary case,
because only with it, a particle number balance can be maitained between the acceleration region 
and the diffusion region.

\subsubsection{Localized acceleration in the center}
\label{refl:center}

\begin{figure*}
\centering
\includegraphics[width=0.49\linewidth]{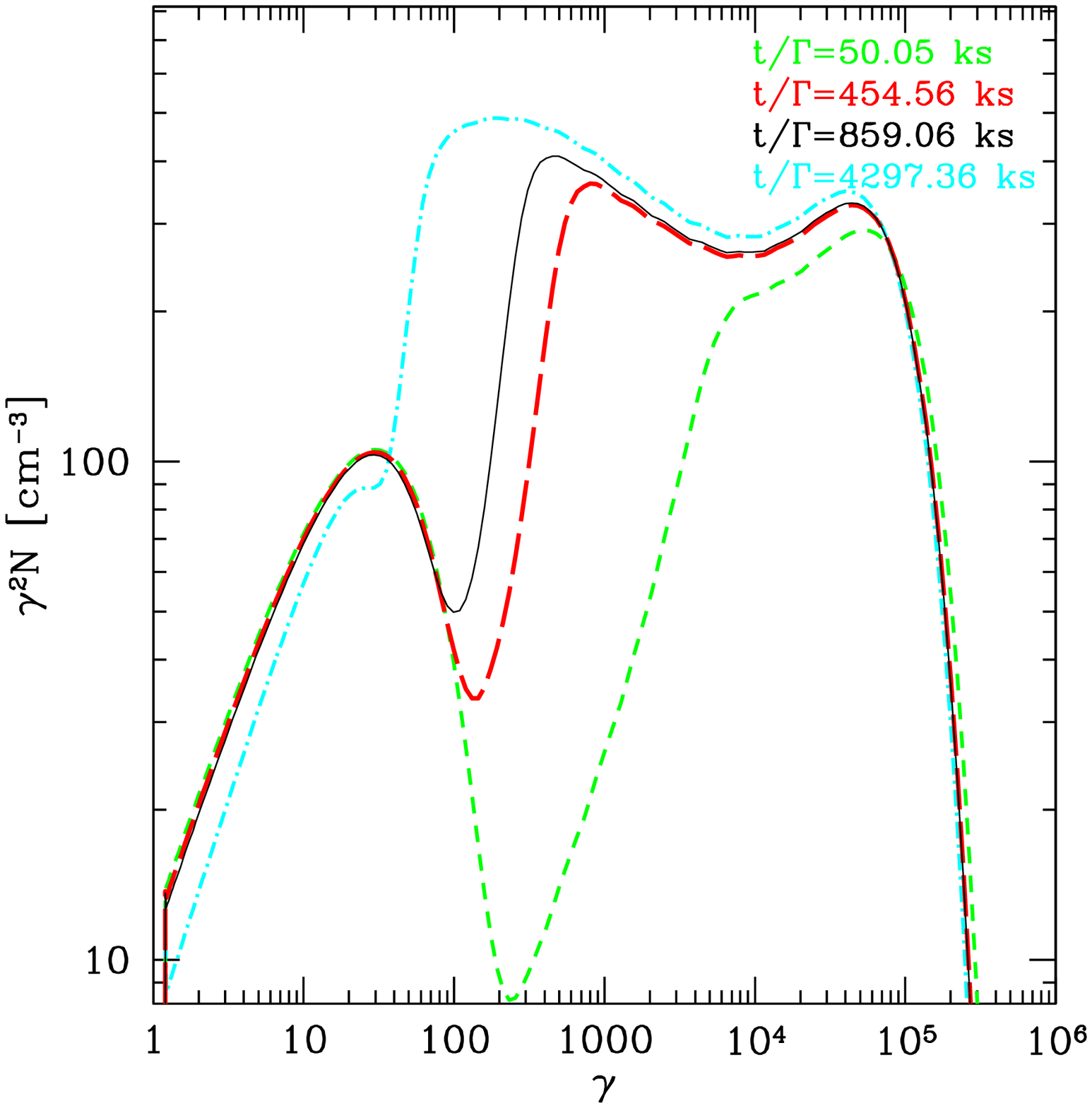}
\includegraphics[width=0.49\linewidth]{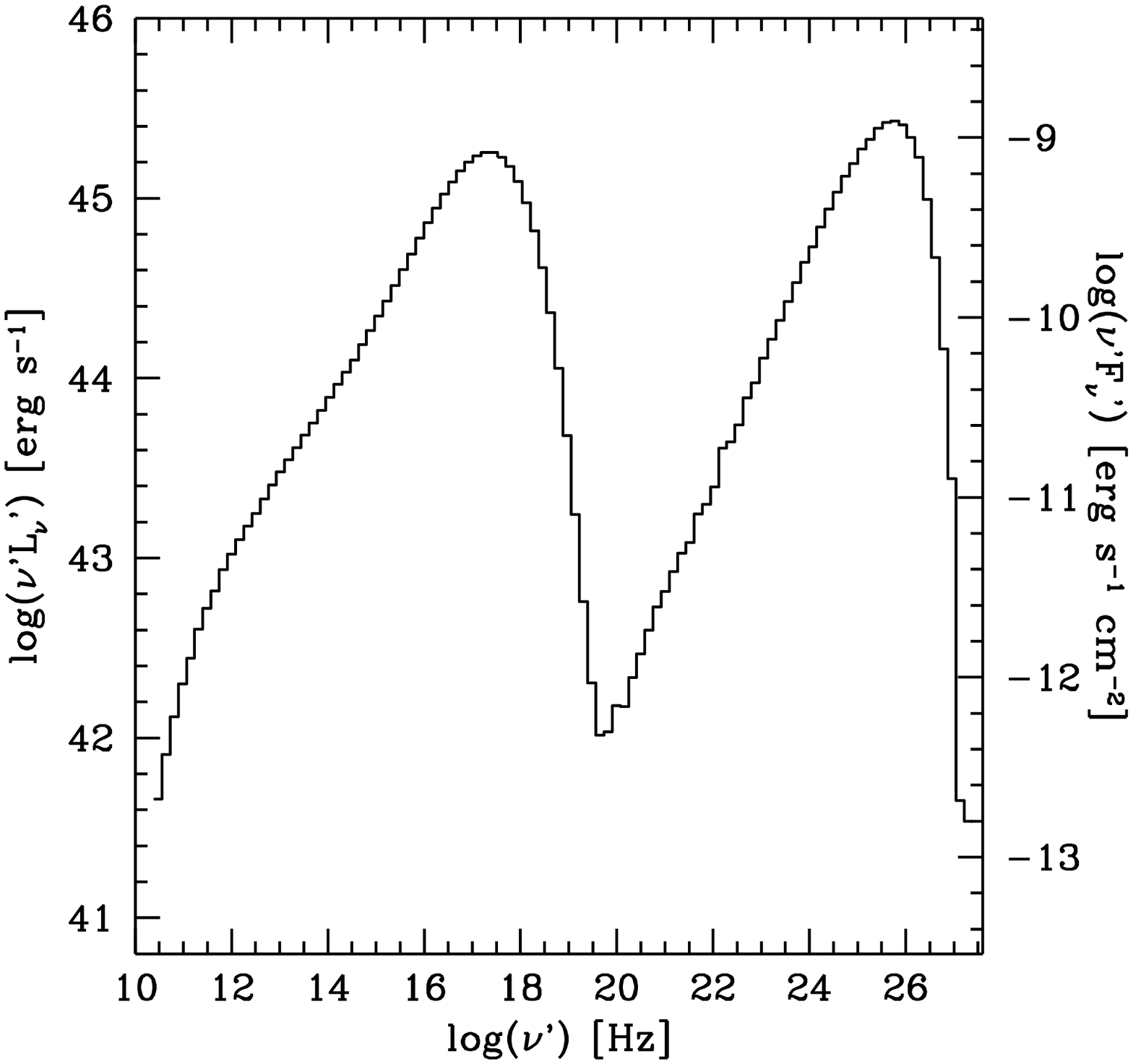}
\includegraphics[width=0.33\linewidth]{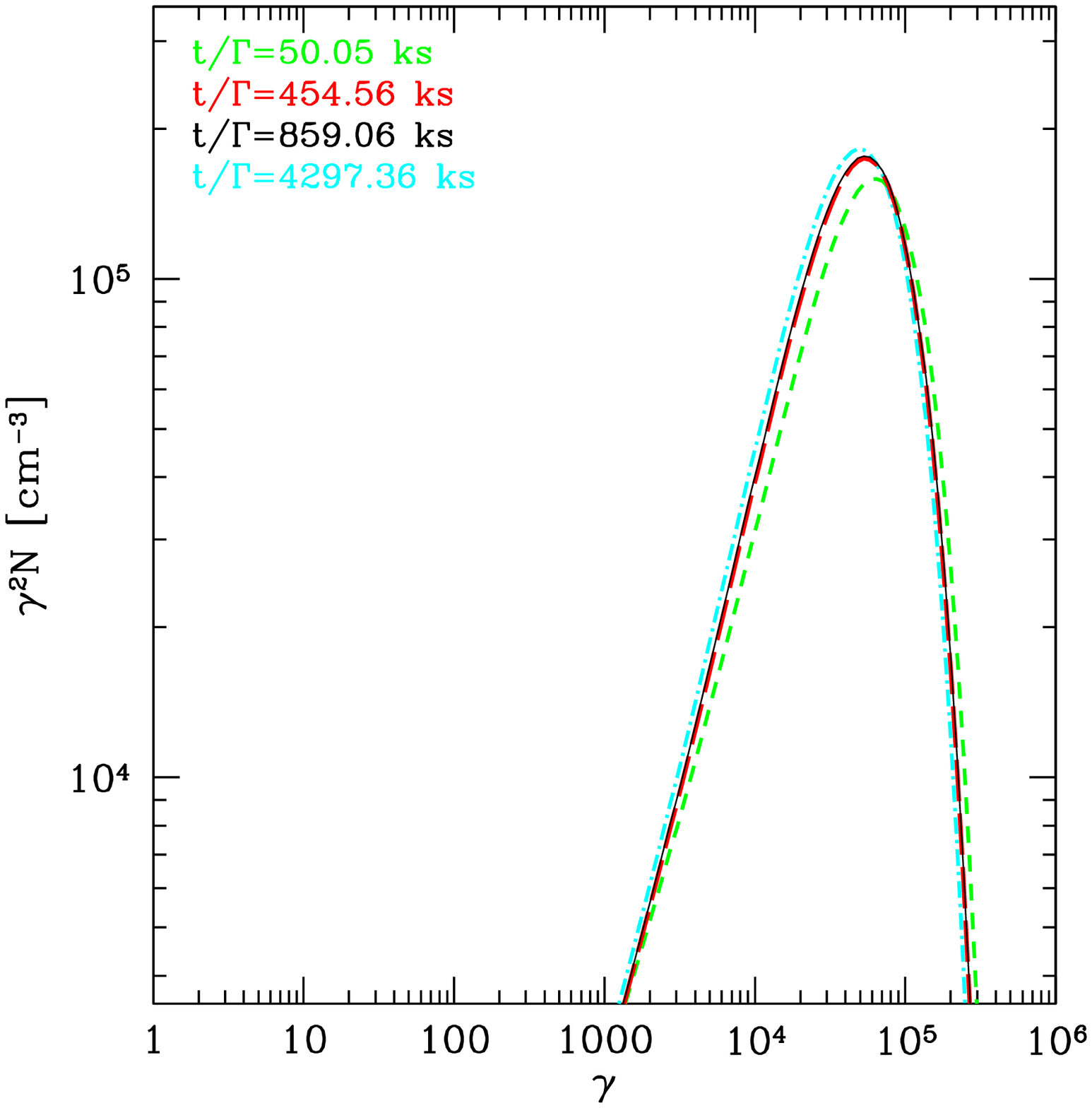}
\includegraphics[width=0.33\linewidth]{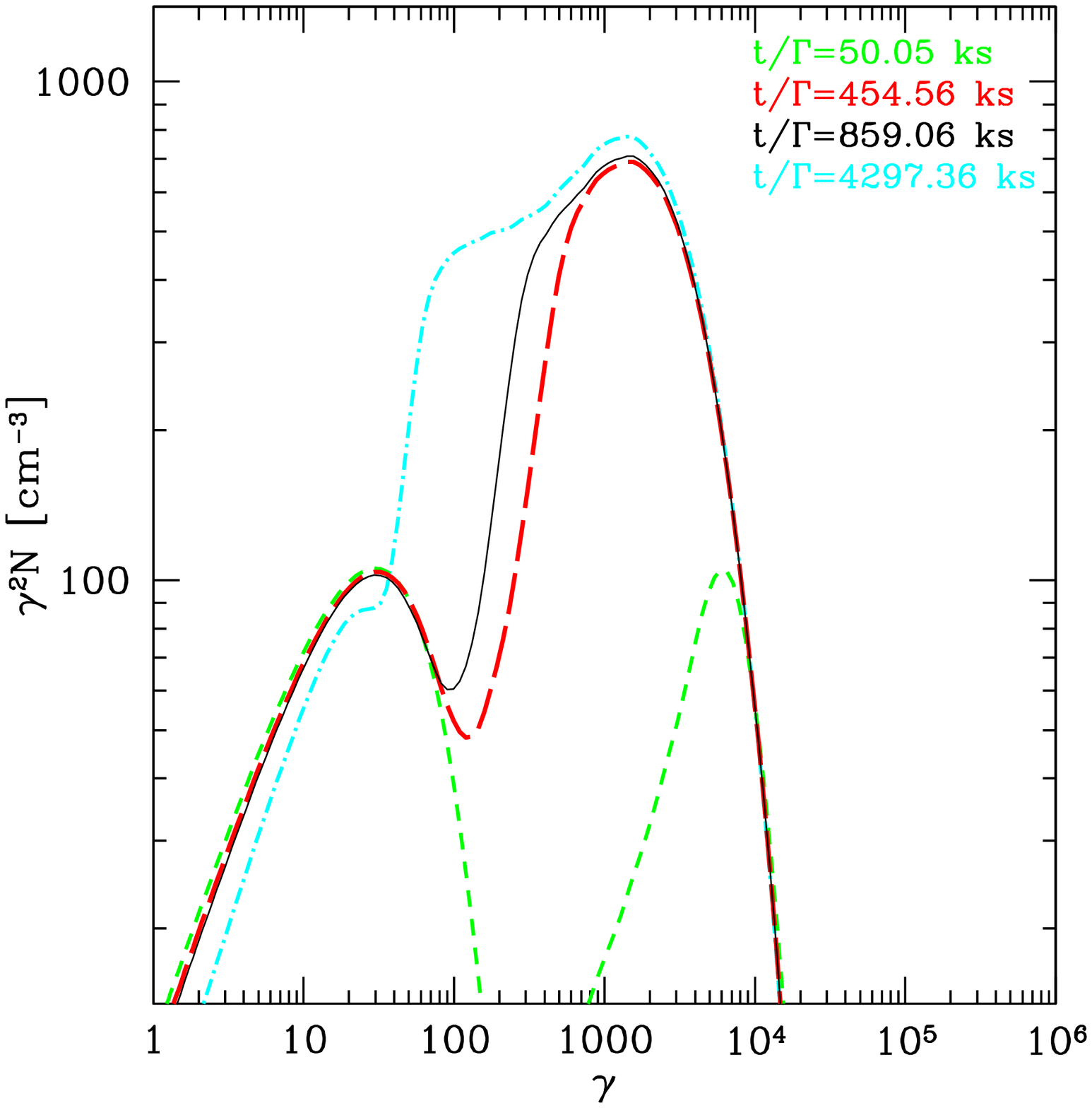}
\includegraphics[width=0.33\linewidth]{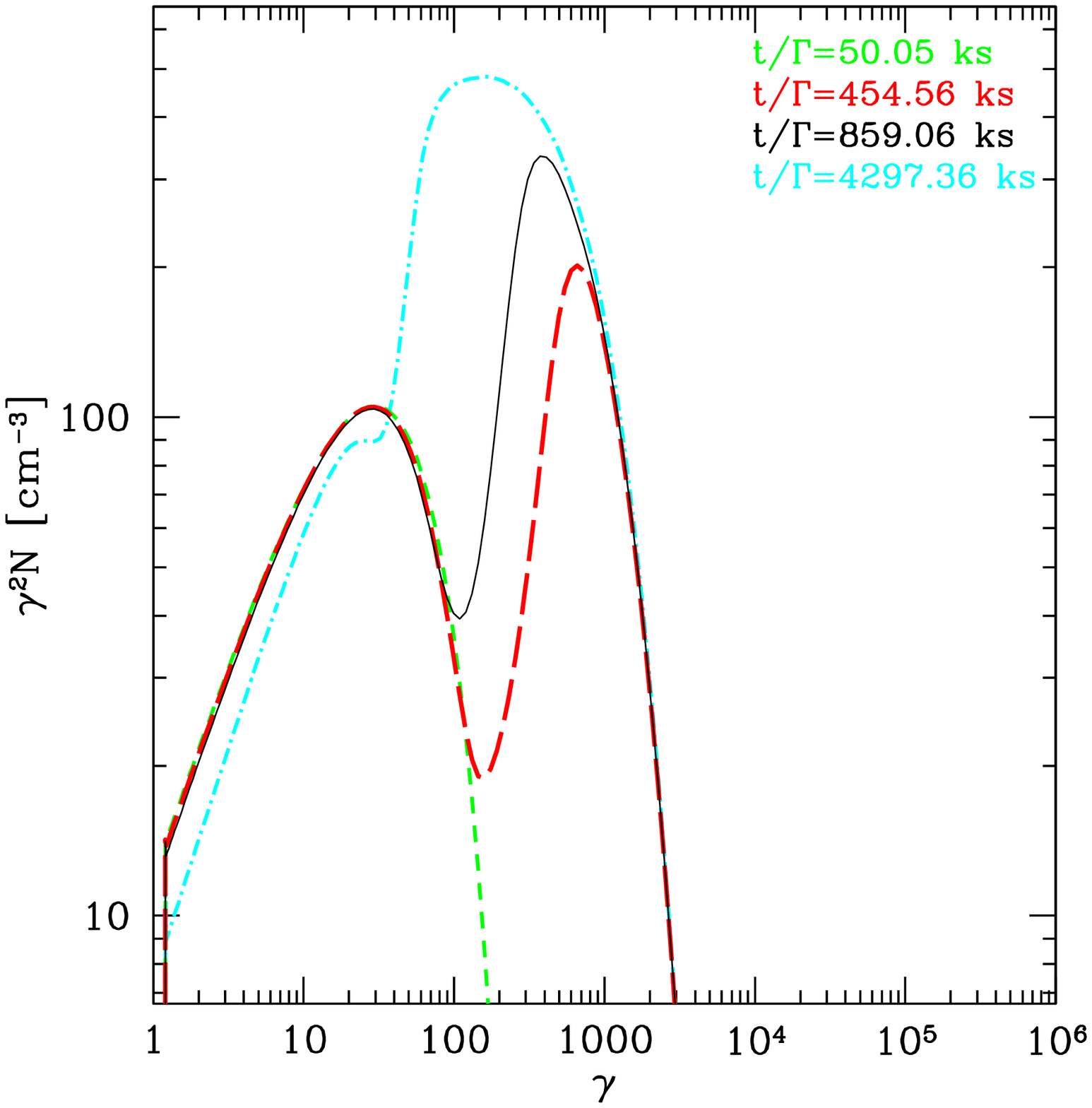}
\caption{(Top)Total EED and SED for the case of \S \ref{refl:slow} (closed boundary, slow diffusion, with 
acceleration in the central region; 
The SED have spectral indices -0.60 at 0.14\,eV, -0.51 at 50\,eV and -0.42 at 10\,GeV.
(Bottom) The EED from the same individual cells as those in Fig.\,\ref{fig:refl_ct_eed}.
}
\label{fig:refl_bump} 
\end{figure*}

In this case, in a central region with 2x2 cells, particles are continuously
accelerated through momentum diffusion. Subsequently the spatial diffusion in both z and r direction
transports the high energy particles through the emission region.
The time-dependent evolution of this process is shown in the electron energy density 
(equivalent to the area covered by a EED plot like those in Fig.\, \ref{fig:2zonect} left) 
maps of Fig.\,\ref{fig:refl_ct_4emaps}.

The cyan line in the EED of Fig.\,\ref{fig:refl_ct_eed} shows the distribution close to the steady state, after a long 
simulation time (8500 time steps). However, to save computer time, in most of the other cases in this work
(except \S \ref{refl:slow}) we
only run the simulation to the point of the black line (1700 time steps). This is enough for our
purpose of showing the difference between cases. For \S \ref{escape}, those
time is already more than enough for the simulation to reach a steady state.

The semi-steady total EED forms a broken power-law distribution where the slope before the break is close to 0.
At early stages, the total EED shows two peaks, because it contains electrons from different regions,
in some of which the particles are accelerated, while in others the particles remain close to their initial distribution.
We also show the SED that comes from the entire volume at late stages.

Single-cell EED from three sample cells are also shown (Fig.\,\ref{fig:refl_ct_eed} bottom). The inner cell
shows that the particles are accelerated to a power-law distribution with very hard spectrum.
The EED with broken power-law distribution in the middle cell (cyan) is a result of the subsequent transport and cooling of those particles.
In the outer cell, the particles had even more time to cool, therefore peak at lower energy compared to the middle cell.

This case is used as the benchmark case for the closed boundary scenario. Main parameters are shown in
Table\,\ref{tab:par}.

\begin{figure*}
\centering
\includegraphics[width=0.49\linewidth]{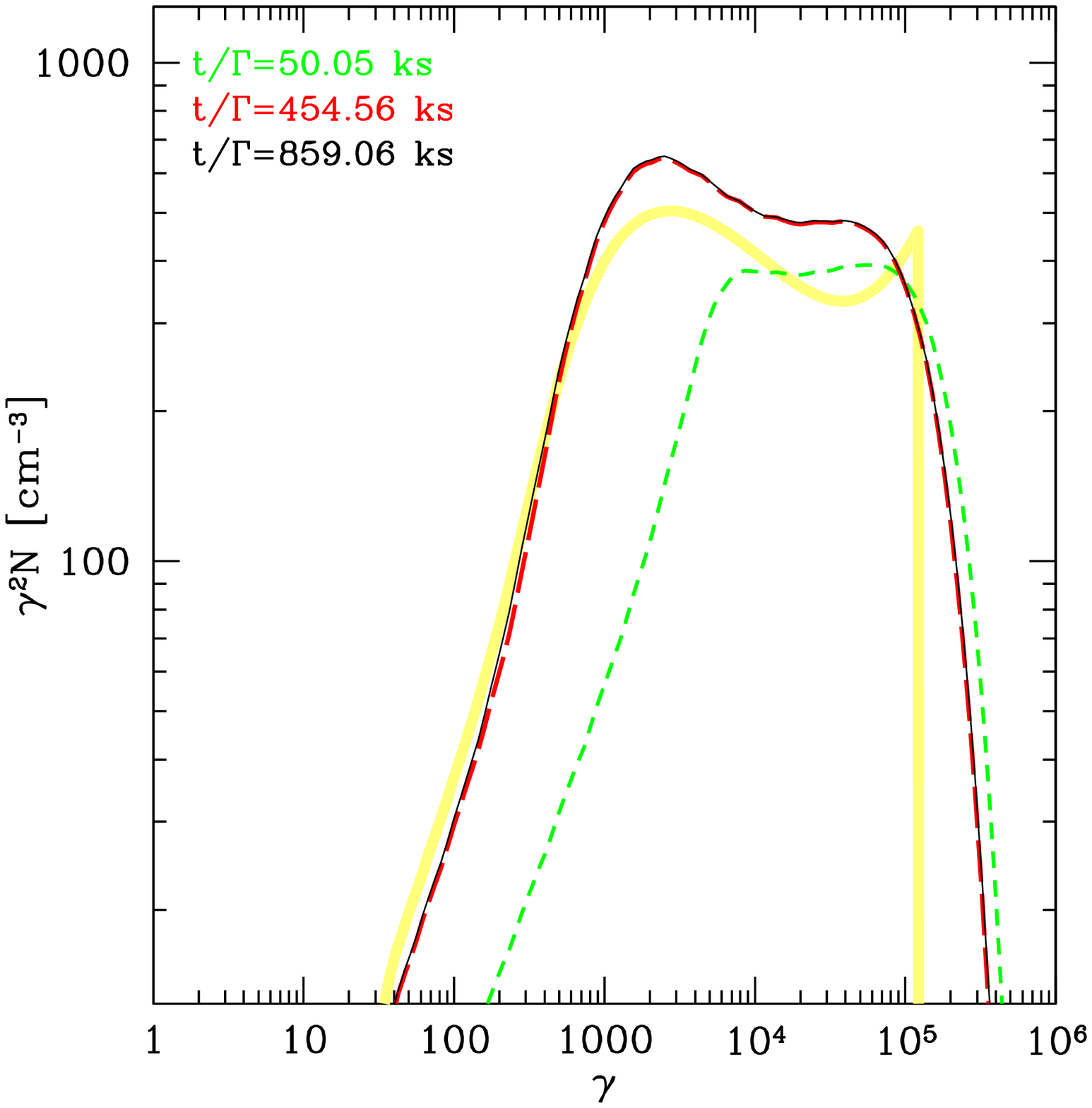}
\includegraphics[width=0.49\linewidth]{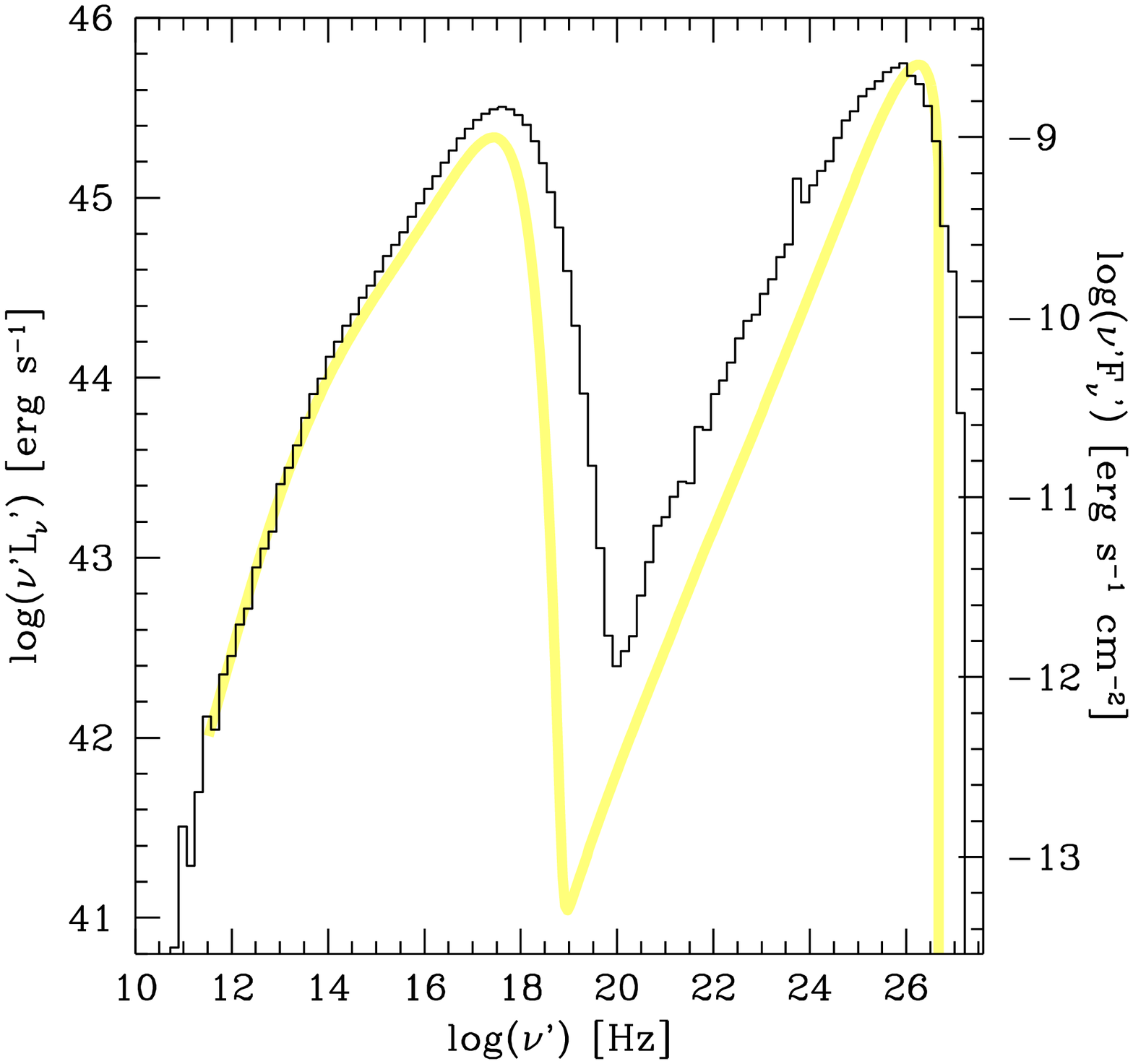}
\caption{Total electron distribution and SED for \S \ref{esc:slow} (open boundary, slow diffusion).
The spectral indices of the SED are -0.57 at 10\,eV, -0.56 at 50\,eV and -0.48 at 10\,GeV
(due to the higher noise in the SED of this case, these estimation is done in a slightly broader frequency rangecompared to the other cases).
The thick yellow lines are the EED and SED from the two-zone model with slow particle escape (Fig.\,\ref{fig:2zonebump}), plotted here for comparison.}
\label{fig:esc_bump}
\end{figure*}

Because the particles are exposed to radiative cooling without further acceleration after they leave the central
acceleration region, the highest-energy particles can only survive in a small central region.
This region gets smaller with increasing particle energy (Fig.\,\ref{fig:refl_ct_maps}).
This energy-dependent jet morphology means that,
by making observation at different frequency,
effectively we may be observing emission region of different size.
The variability at different energy may still be correlated, but there might be a significant difference
in the light curves. More details of the variability pattern of the jet with localized acceleration will 
be discussed in a separate publication.
The energy dependence also affects the SSC scattering. As we have already discussed in \S,\ref{2zone},
the concentration of the most energetic photons and electrons in the center causes the SSC spectrum to
be harder than the synchrotron spectrum. This feature is clearly seen in
both the two-zone (Fig.\,\ref{fig:2zonebump} right) and 
the 2D (Fig.\,\ref{fig:esc_bump} upper-right) models, even though the confined diffusion
scenario is quite different from the two-zone model.
But it would not be expected in a one-zone model.
This energy dependence also implies that even though the highest-energy photons are produced in a
very compact region, the lower-energy photons, which may cause pair creation with the
high energy photons, are less concentrated, thus alleviating the compactness constraint 
\citep{boutelier_2008:inhom:390.73}.

\begin{figure*}
\centering
\includegraphics[width=0.49\linewidth]{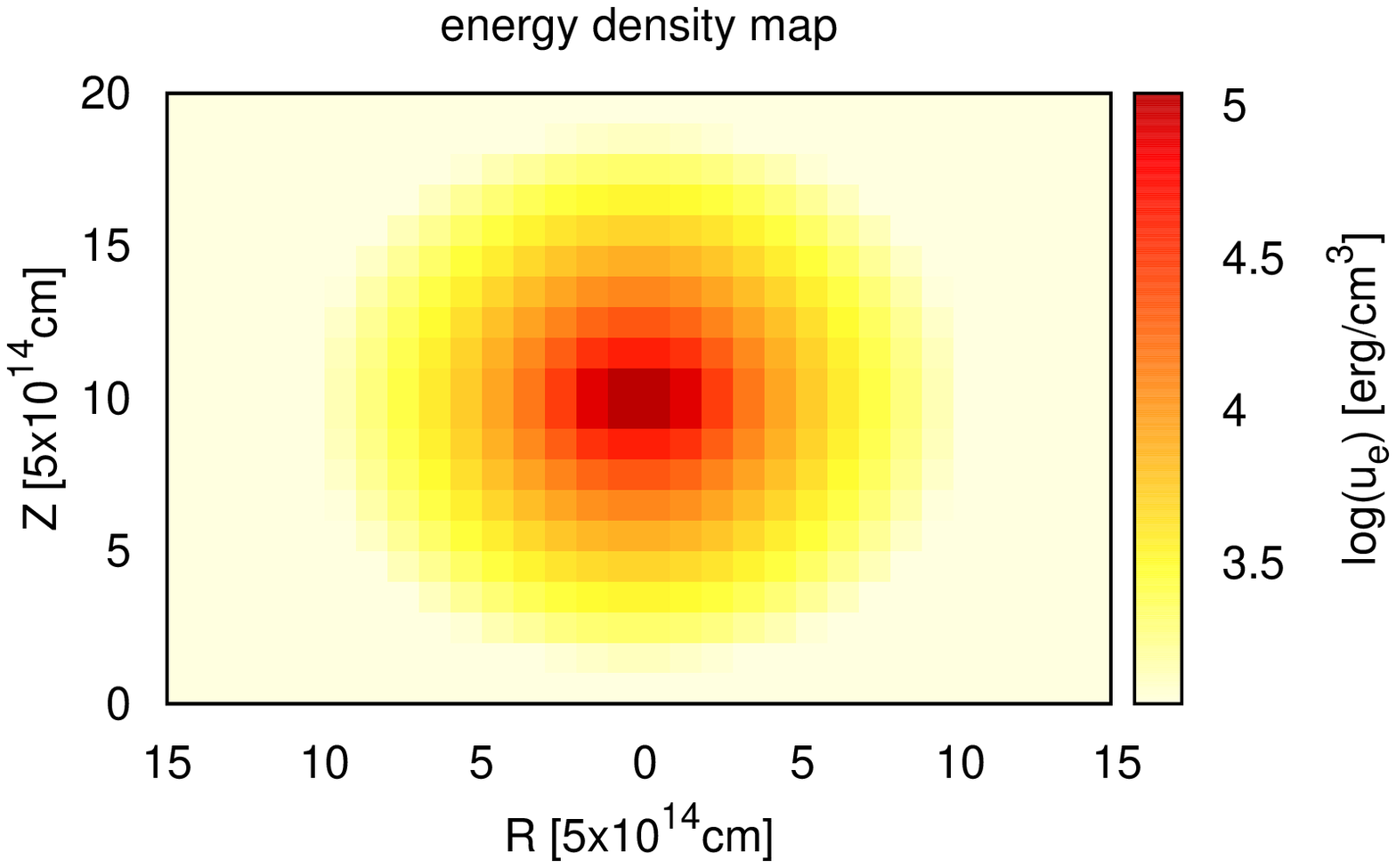}
\includegraphics[width=0.49\linewidth]{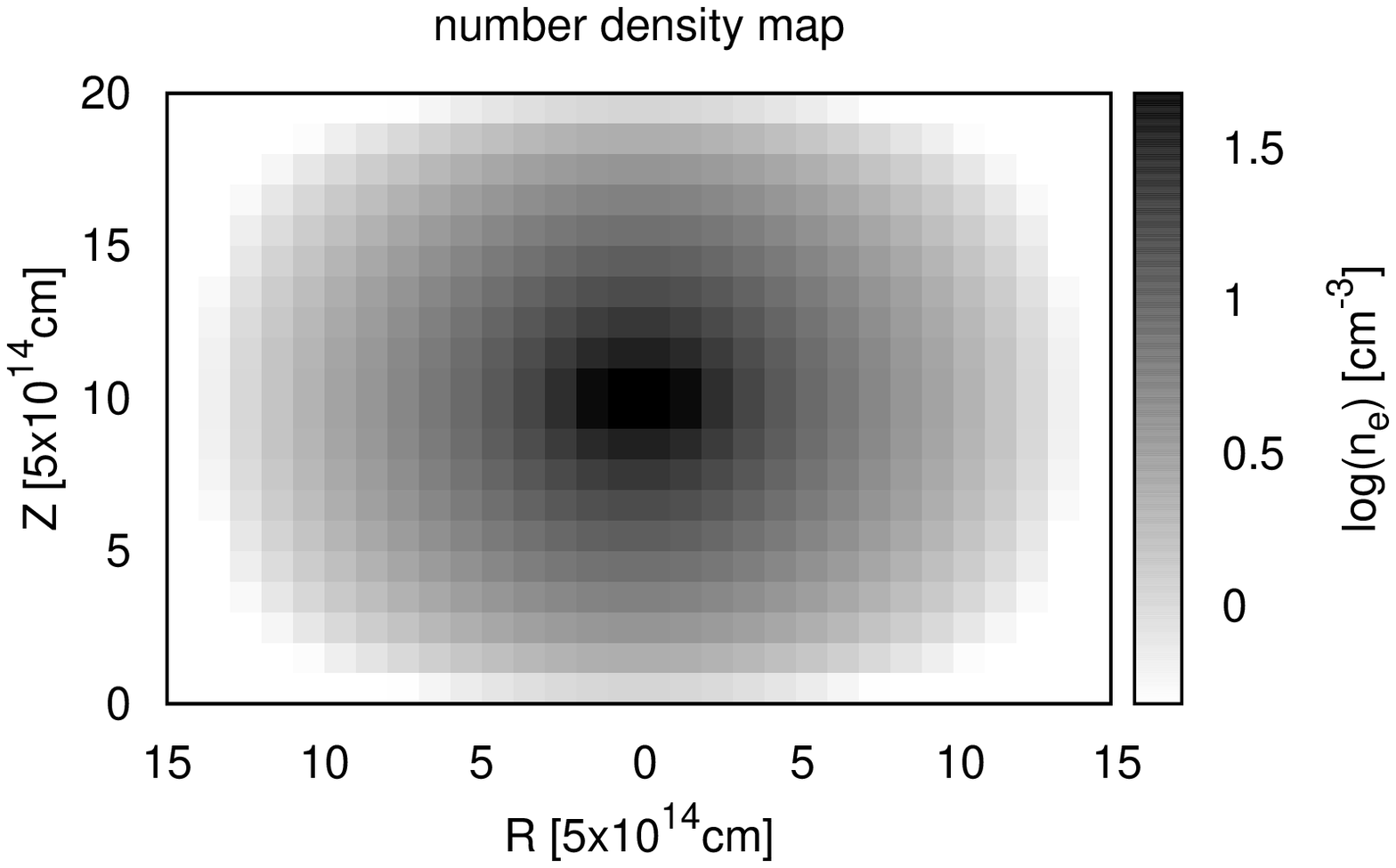}
\caption{Density maps for the case of \S \ref{esc:center} (open boundary, acceleration in the center) at simulation time step 1700.
}
\label{fig:esc_ct_maps} 
\end{figure*}

\begin{figure*}
\centering
\includegraphics[width=0.49\linewidth]{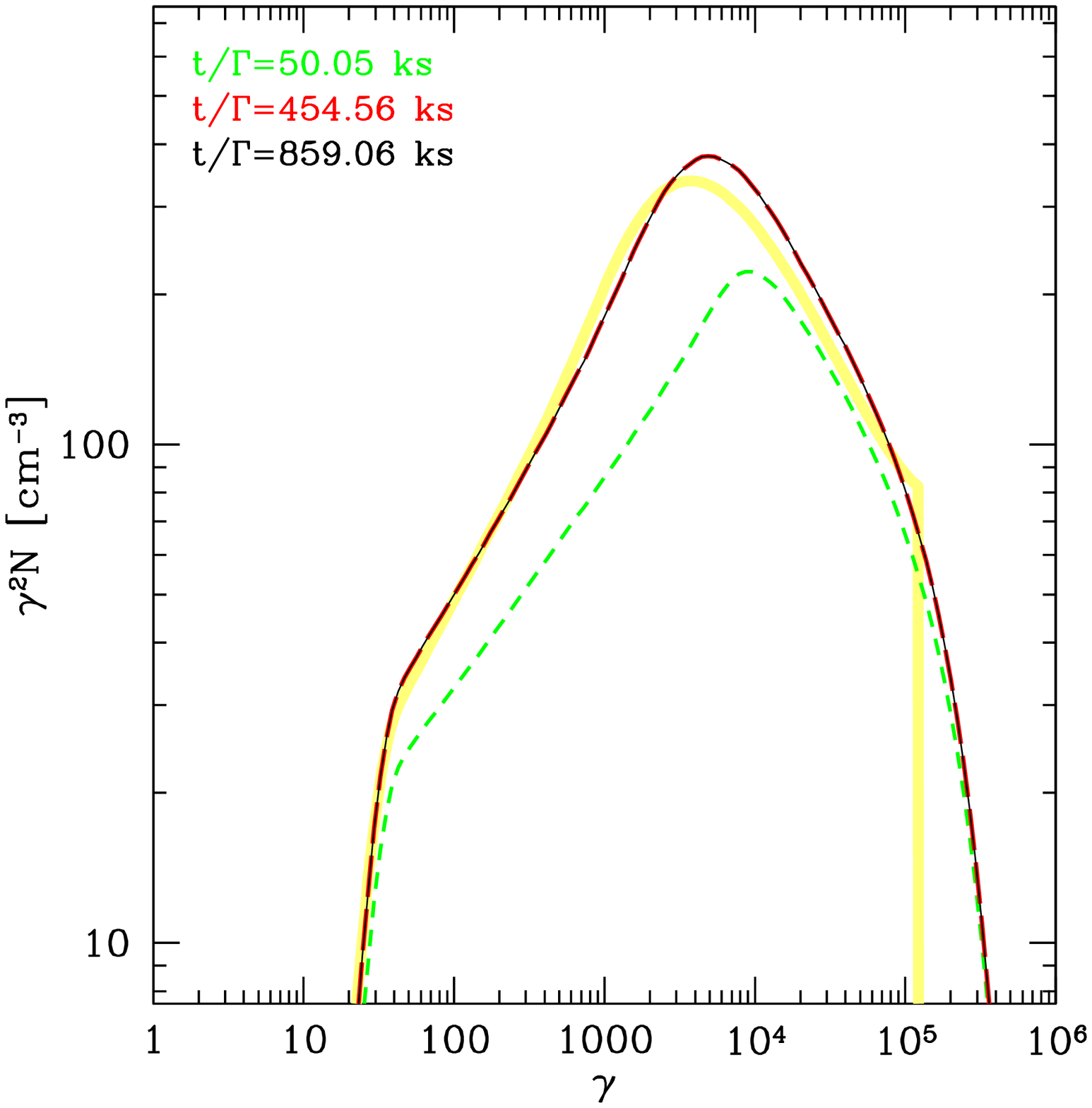}
\includegraphics[width=0.49\linewidth]{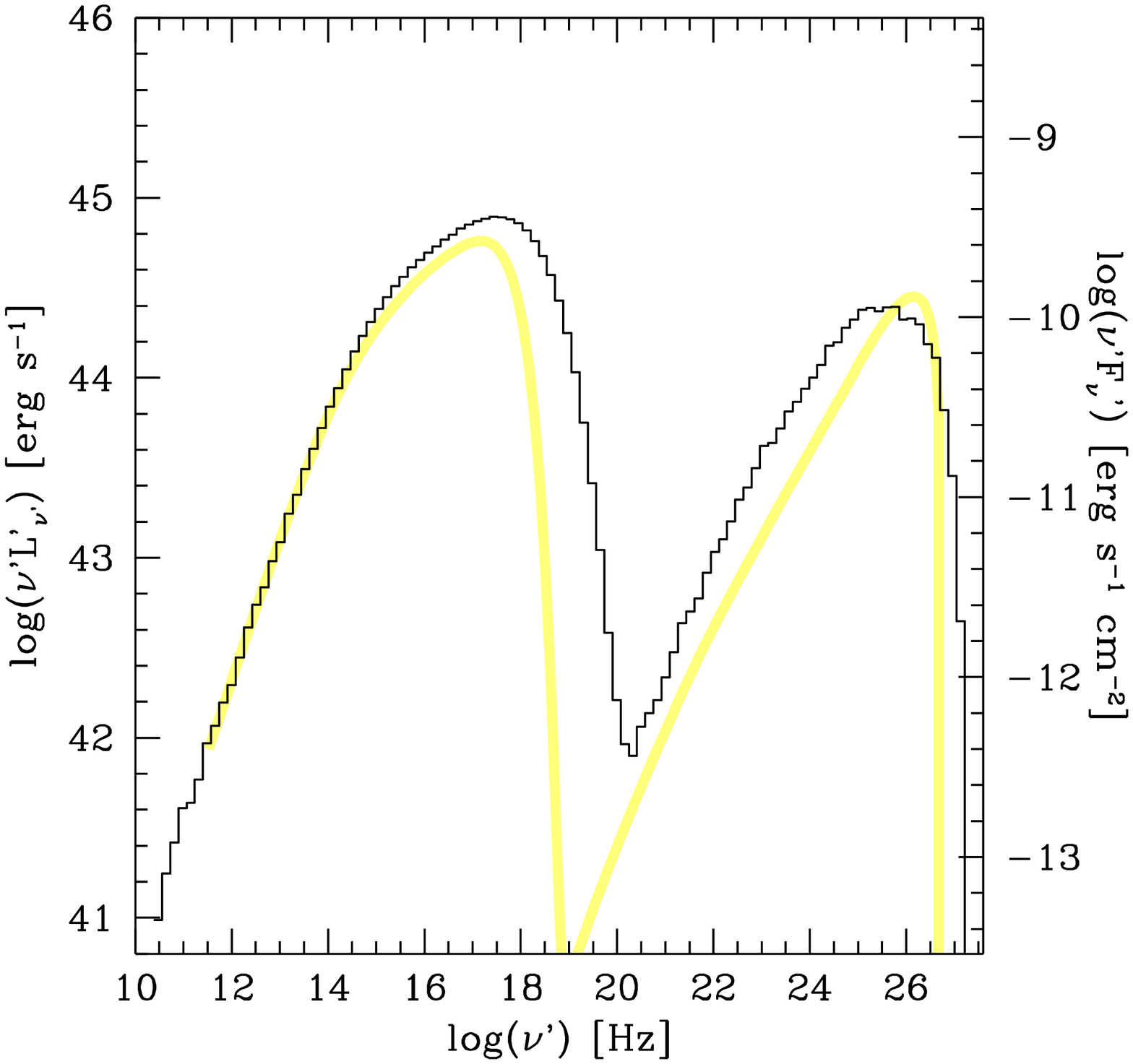}
\includegraphics[width=0.33\linewidth]{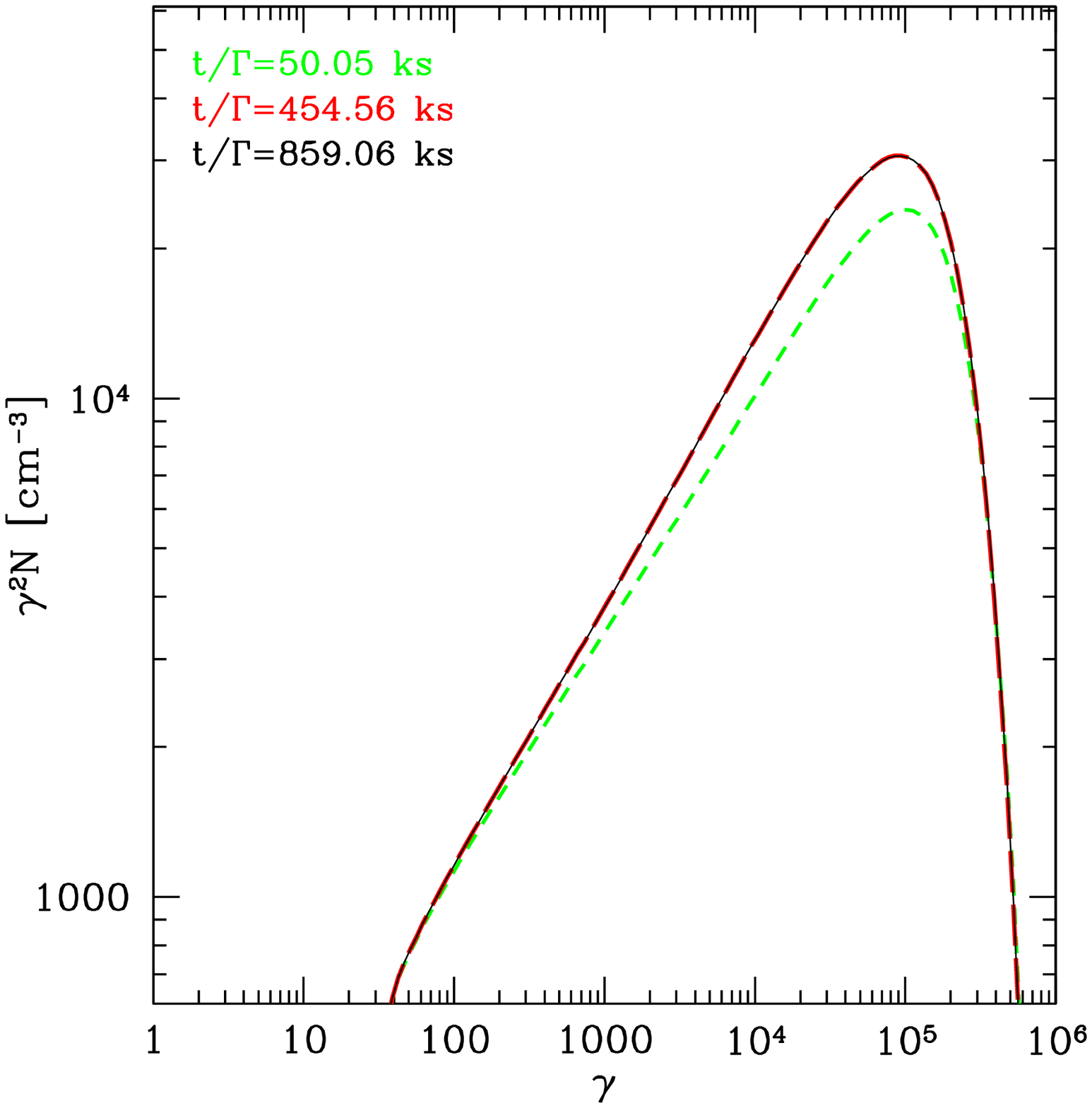}
\includegraphics[width=0.33\linewidth]{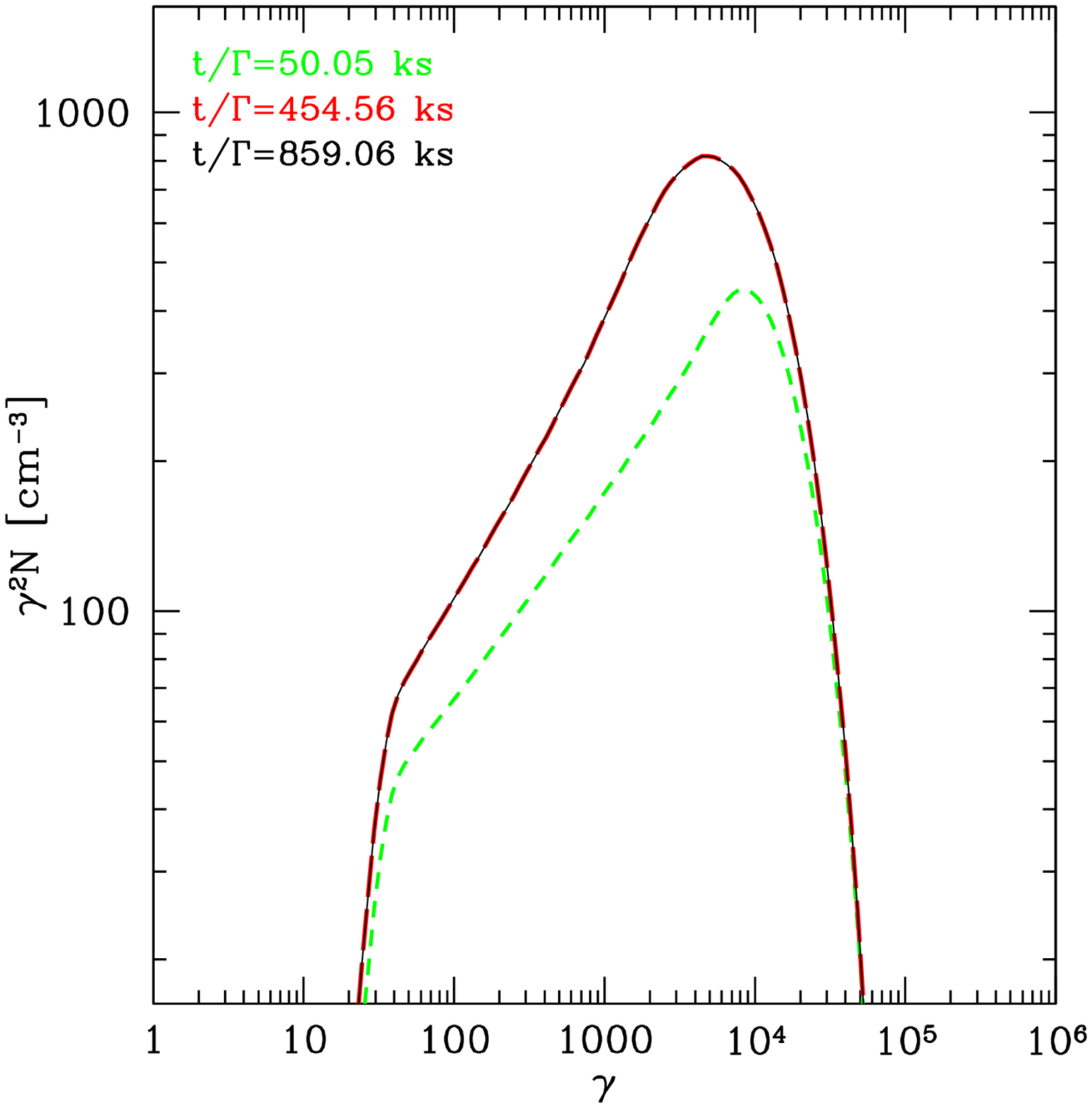}
\includegraphics[width=0.33\linewidth]{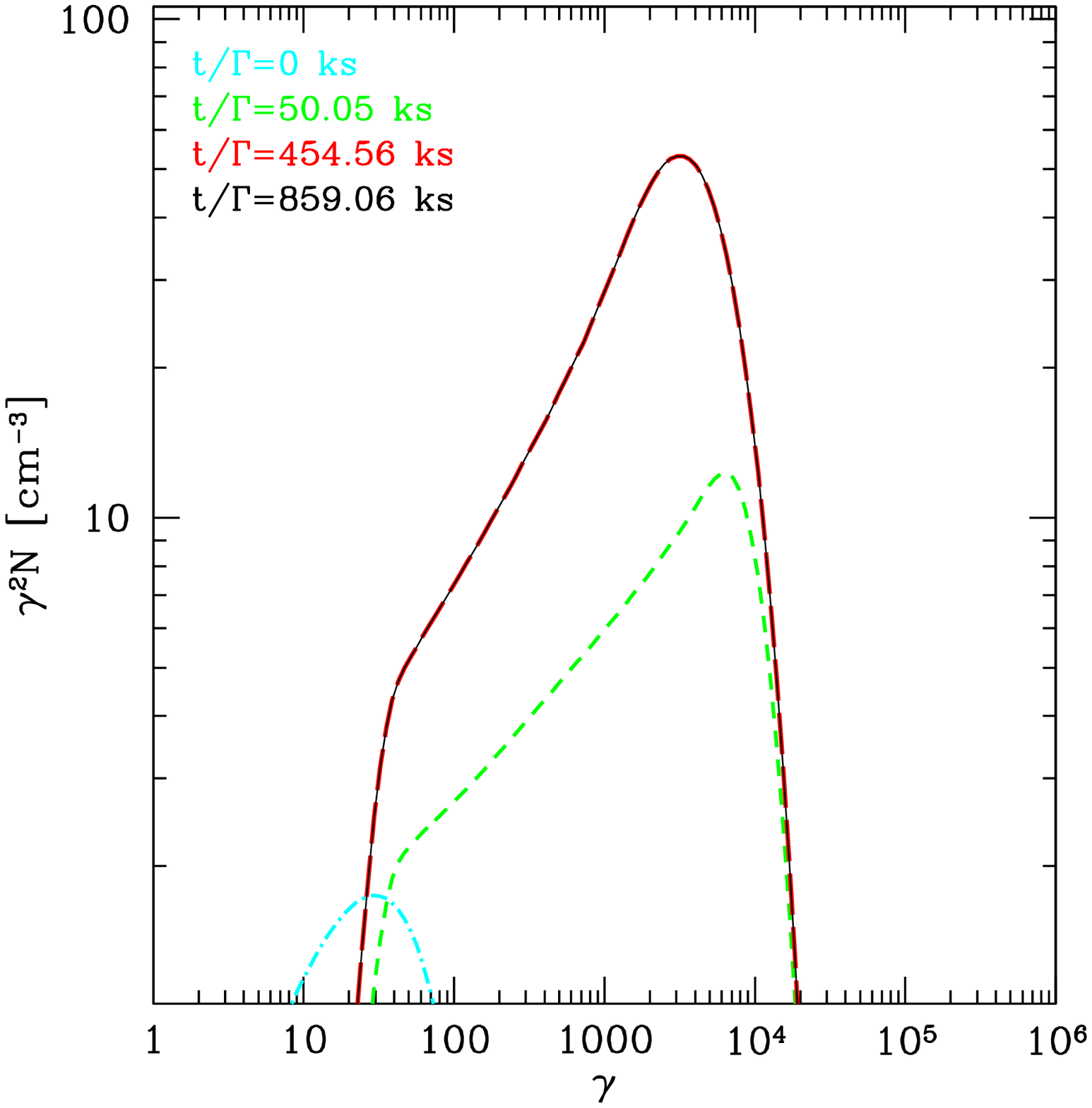}
\caption{(Top) Total EED and SED for \S \ref{esc:center} (open boundary, acceleration in the center). 
The thick yellow lines are the EED and SED from the two-zone model (Fig.\,\ref{fig:2zonect}), plotted here
for comparison.
The spectral indices of the EED are -1.50 at $\gamma=2\times10^2$ and -2.52 at $\gamma=2\times10^4$.
The spectral indices of the SED are -0.63 at 10\,eV and -0.53 at 1\,GeV.
(Bottom) The EED from the same individual cells as those in Fig.\,\ref{fig:refl_ct_eed}.
}
\label{fig:esc_ct_eed}      
\end{figure*}

\subsubsection{Slow diffusion}
\label{refl:slow}

In this case, the particle diffusion is less efficient compared to \S \ref{refl:center}.
This results in slower rate of particle escape from the acceleration region, and therefore
harder EED (lower-left of Fig.\,\ref{fig:refl_bump}).
Similar to the two-zone model with slow particle escape (Fig.\,\ref{fig:2zonebump}),
the particles in the acceleration region
have excess energy that provides an additional bump at the high-energy end of the total EED before cut off.
The \gray spectrum is also extremely hard, with a spectral index of about -0.4 
(equivalent to a photon index of -1.4) at 10\,GeV.
This kind of spectrum generally applies to localized acceleration with slow particle escape
($t_\mathrm{esc,acc}\gg t_\mathrm{acc}$).
The spectral hardening is another consequence of the energy-dependent inhomogeneity we show in 
Fig.\,\ref{fig:refl_ct_maps}. 
For slow particle diffusion, in which the high-energy bump is apparent,
the power-law slope before the bump should always be close to 2, because it is the radiatively-cooled 
version of the $p\sim1$ spectrum resulted from inefficient particle escape.

\begin{figure*}
 \centering
 \begin{minipage}{0.4\linewidth}
   \includegraphics[width=1.\linewidth]{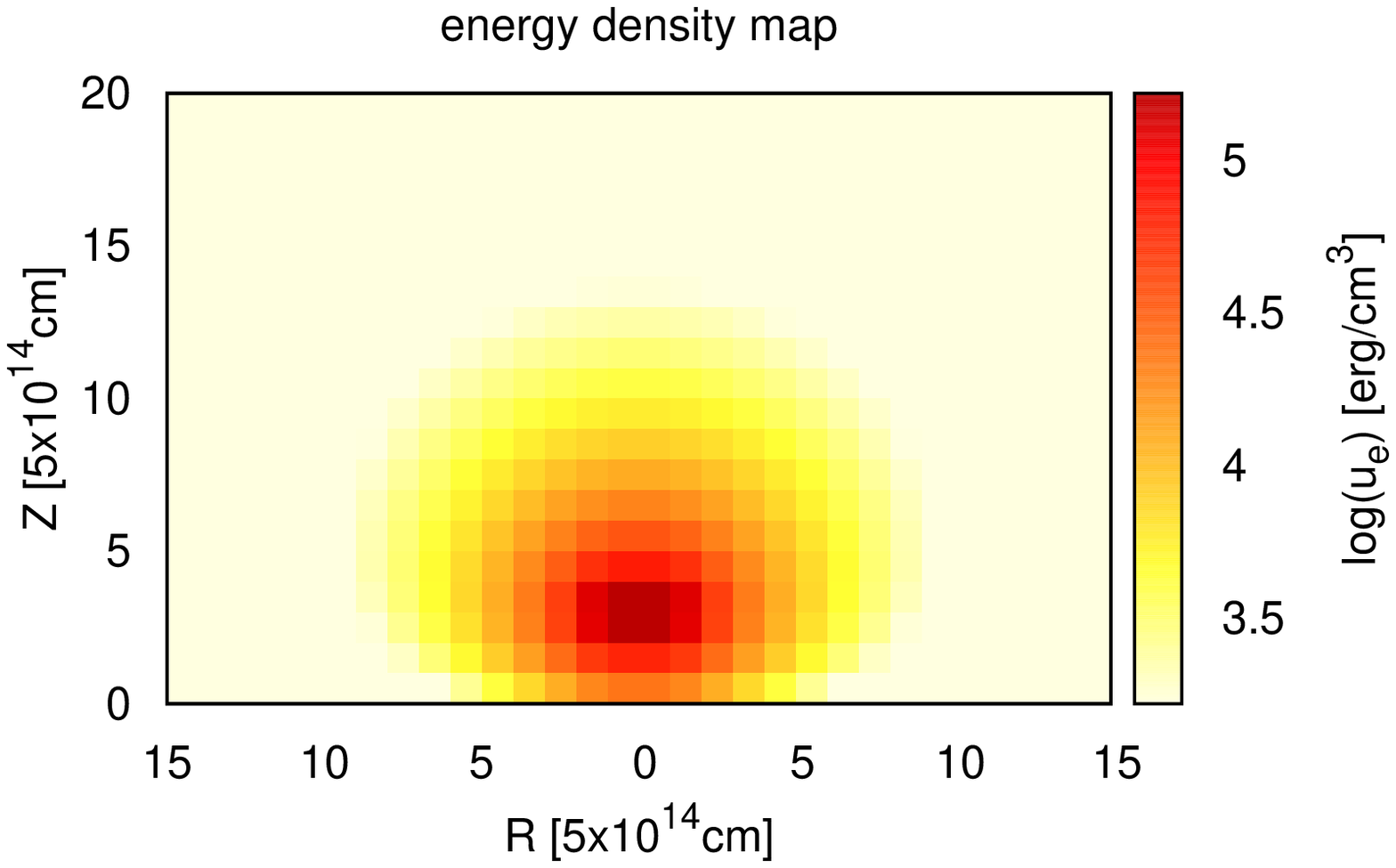}
   \includegraphics[width=1.\linewidth]{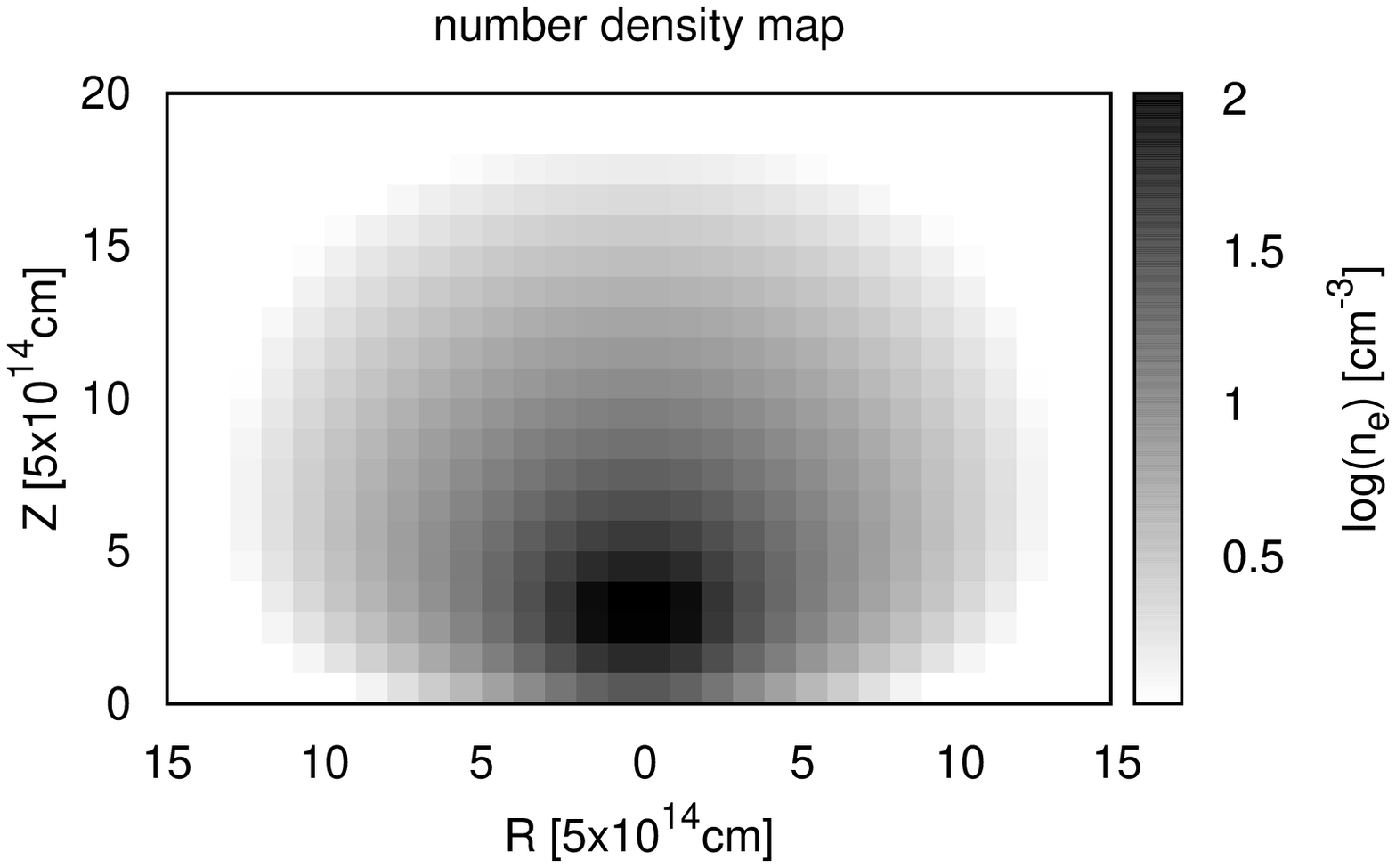}
 \end{minipage}
 \begin{minipage}{0.59\linewidth}
   \includegraphics[width=1.\linewidth]{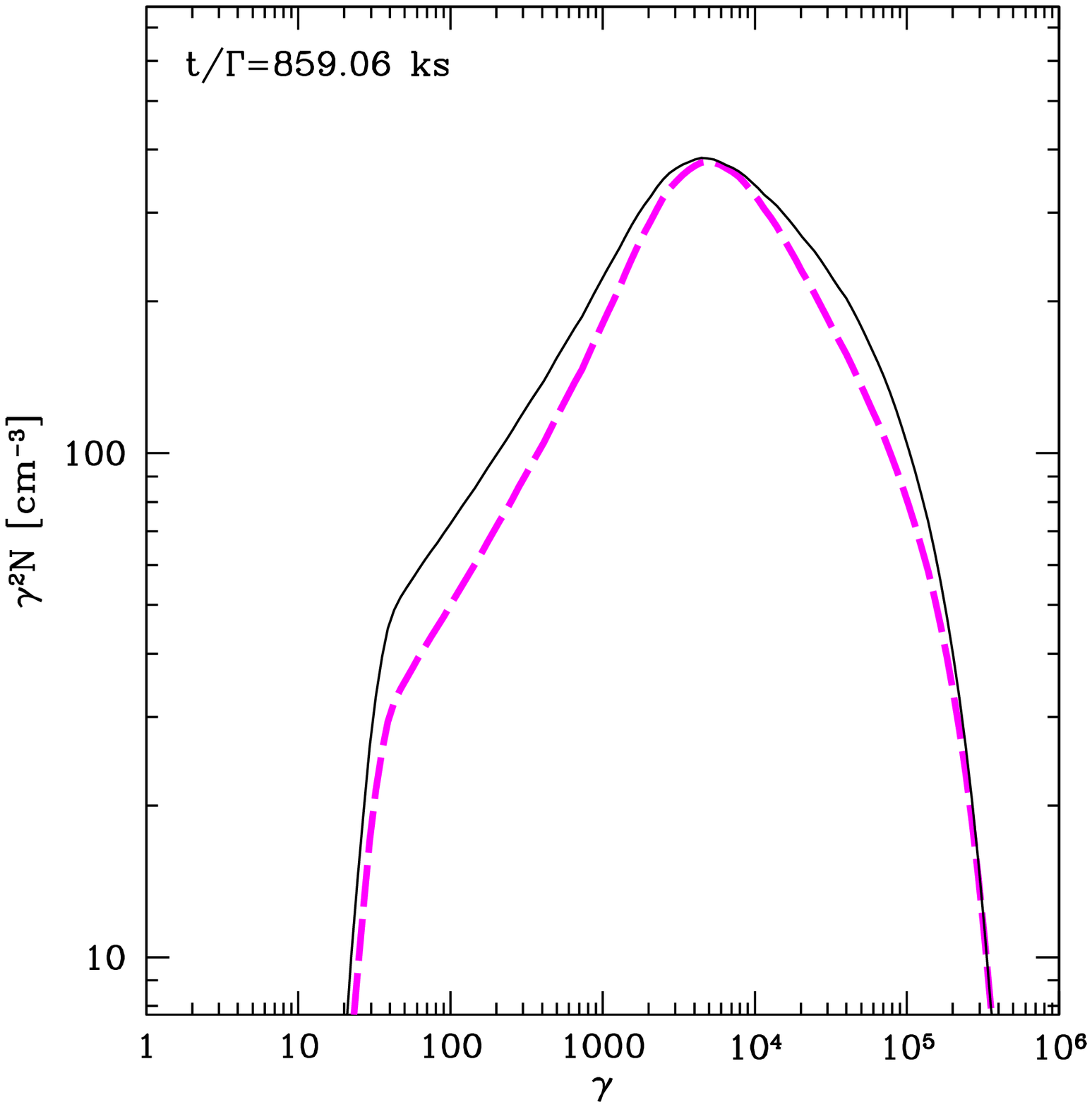}
 \end{minipage}
 \caption{Density maps and total EED (black solid line) for the case of \S \ref{esc:away} (open boundary, acceleration away from the center) at simulation time step 1700. The spectral indices of the EED are -1.57 at $\gamma=2\times10^2$ and -2.34 at $\gamma=2\times10^4$.
The final EED shown in 
Fig.\,\ref{fig:esc_ct_eed} is plotted here for comparison (magenta dashed line).
}
 \label{fig:esc_away} 
\end{figure*}

\begin{figure*}
\centering
 \begin{minipage}{0.4\linewidth}
   \includegraphics[width=1.\linewidth]{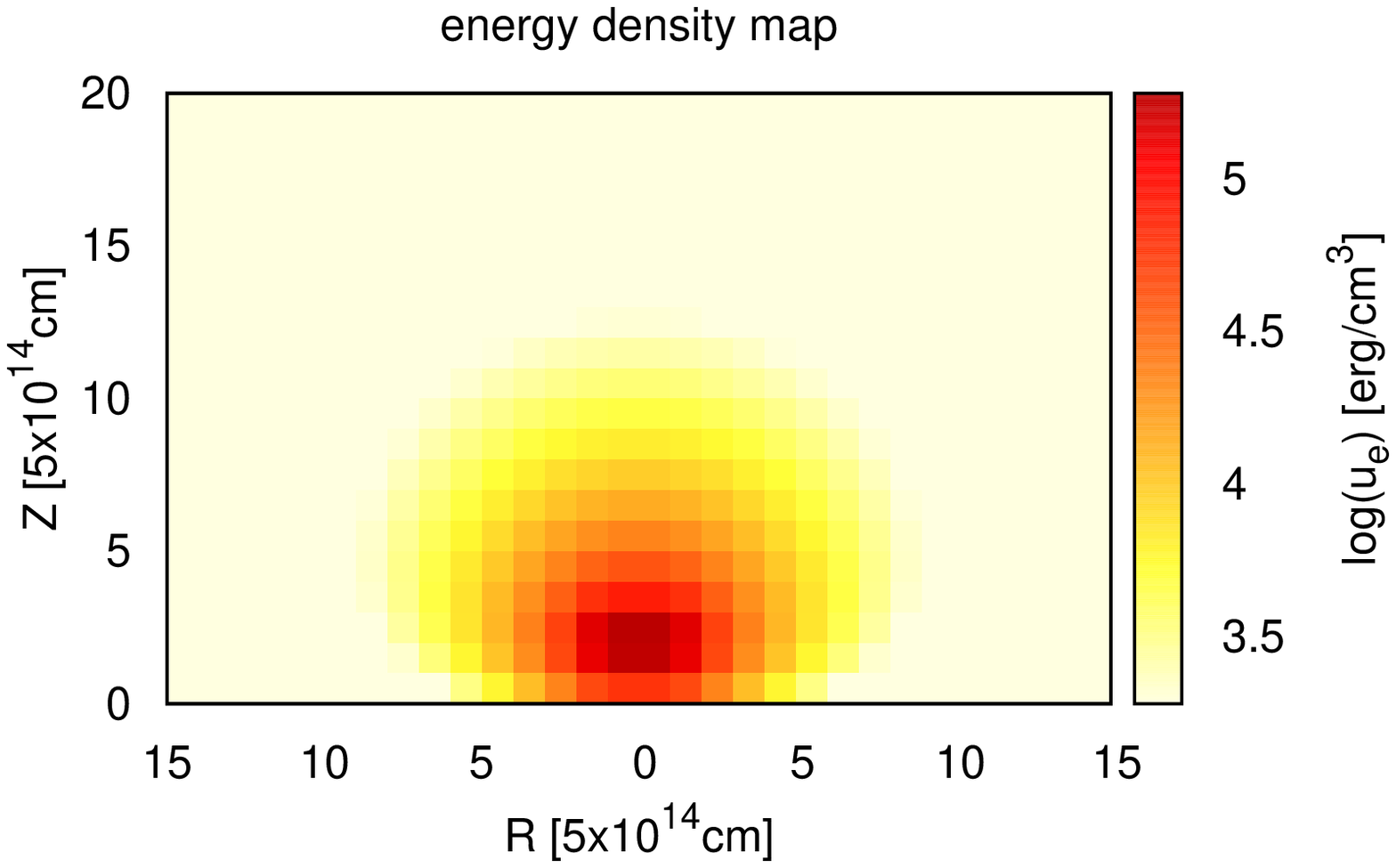}
   \includegraphics[width=1.\linewidth]{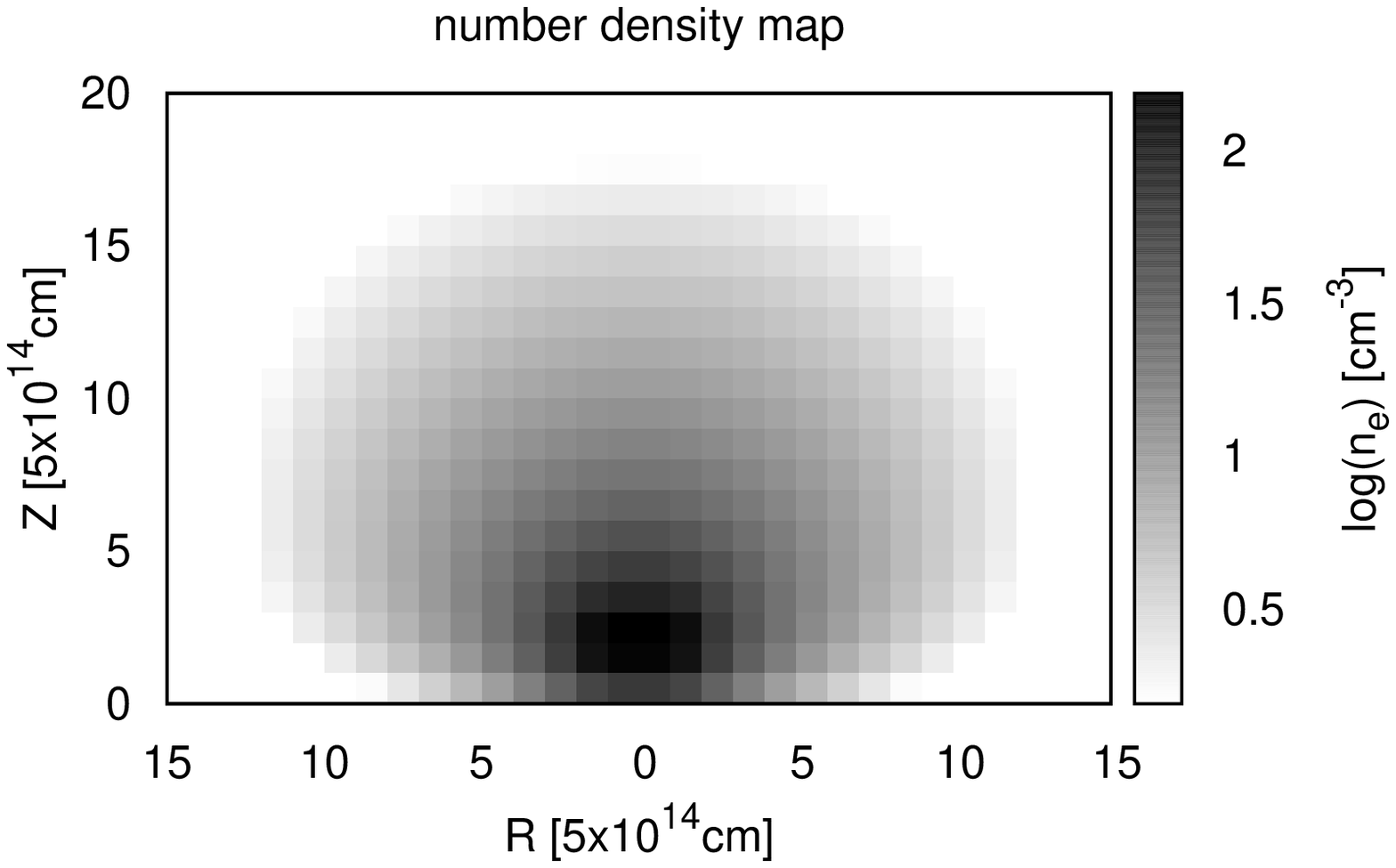}
 \end{minipage}
 \begin{minipage}{0.59\linewidth}
   \includegraphics[width=1.\linewidth]{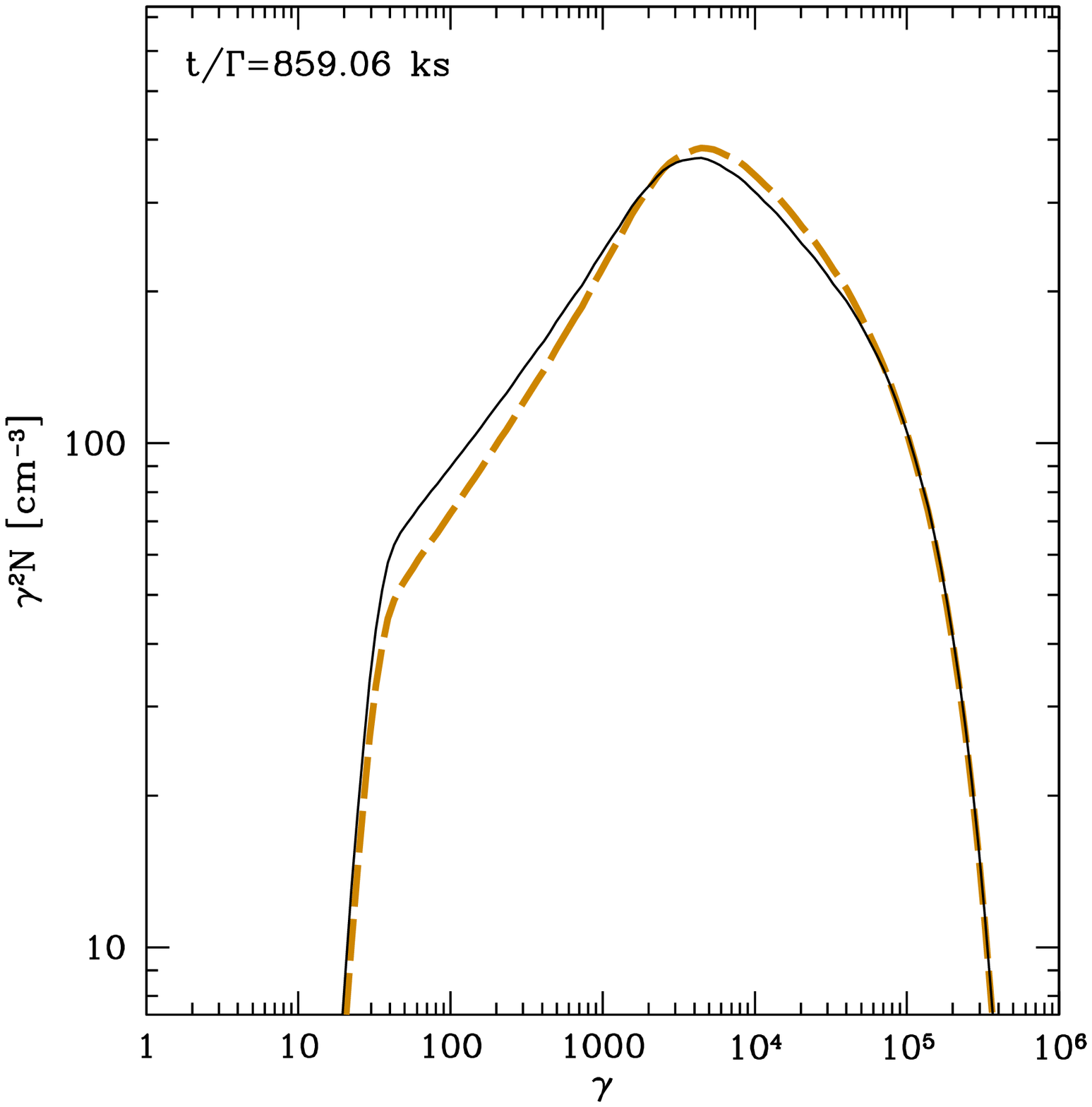}
 \end{minipage}
\caption{Density maps and total EED (black solid line) for the case of \S \ref{esc:away} (open boundary, acceleration away from the center)
at simulation time step 1700. The acceleration region is futher away from the center compared to 
Fig. \ref{fig:esc_away}.
The spectral indices of the EED are -1.62 at $\gamma=2\times10^2$ and -2.33 at $\gamma=2\times10^4$.
The EED shown in 
Fig. \ref{fig:esc_away} is plotted here for comparison (orange dashed line).
}
\label{fig:esc_away2} 
\end{figure*}

\begin{figure*}
\centering
\includegraphics[width=0.49\linewidth]{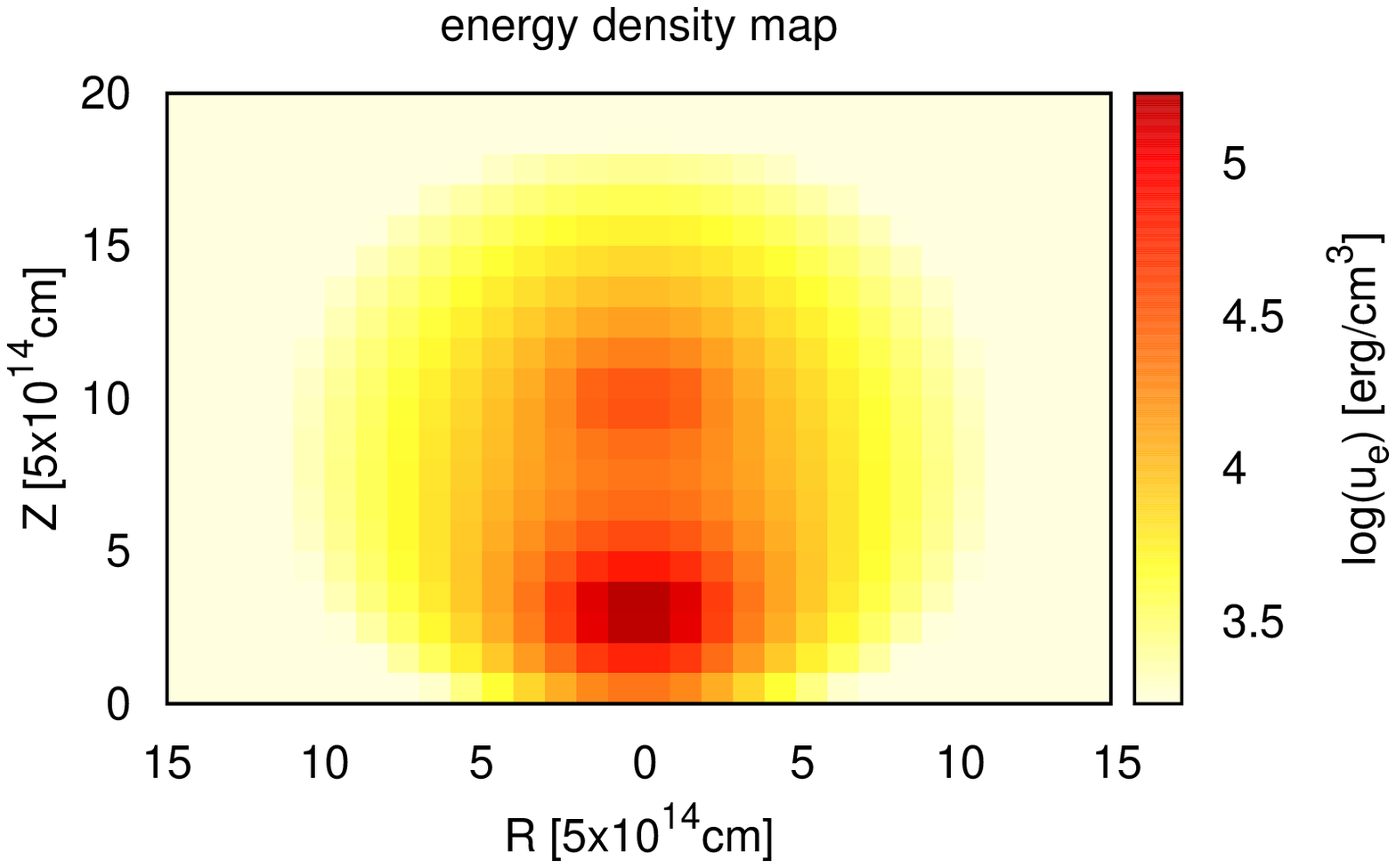}
\includegraphics[width=0.49\linewidth]{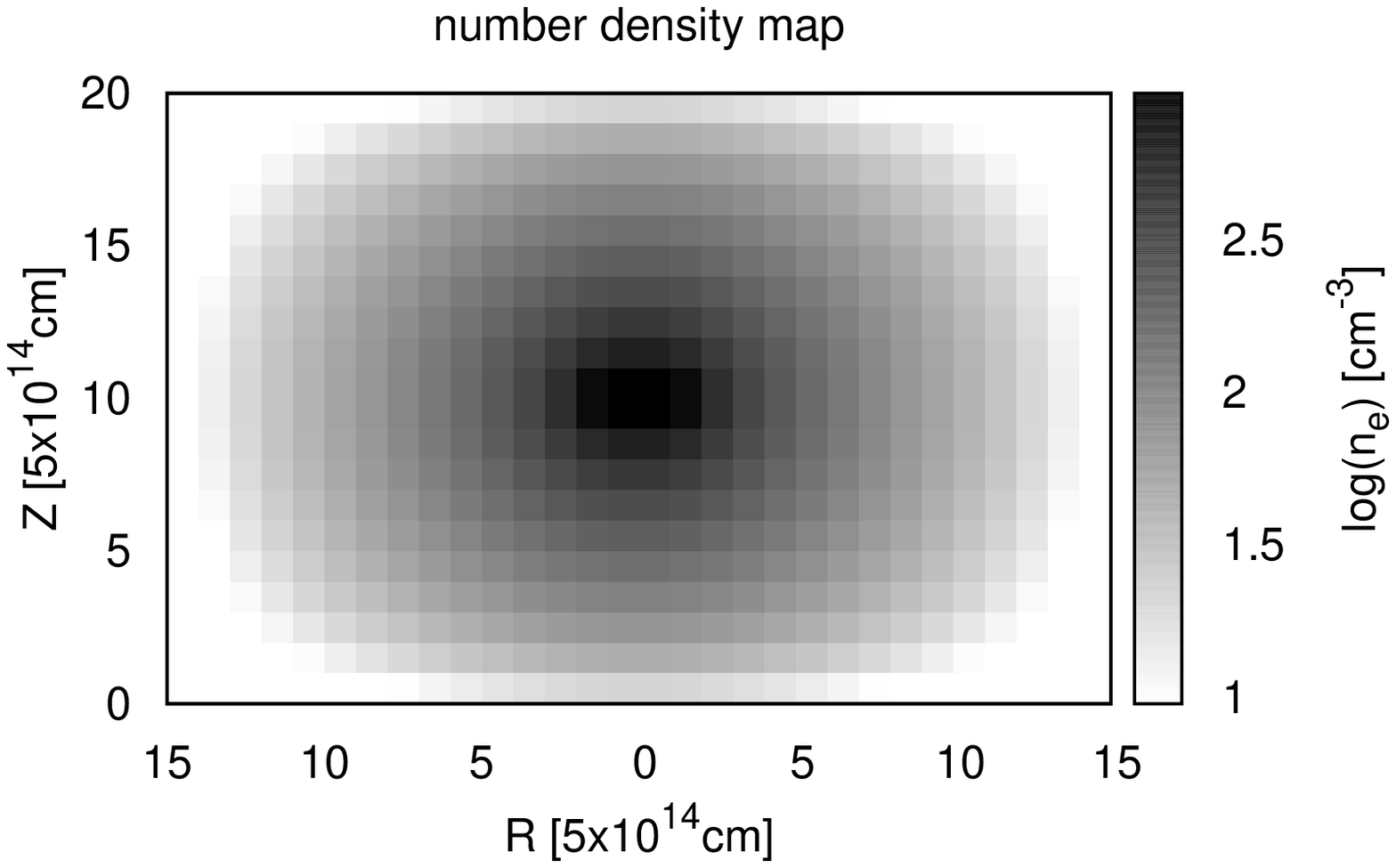}
\includegraphics[width=0.49\linewidth]{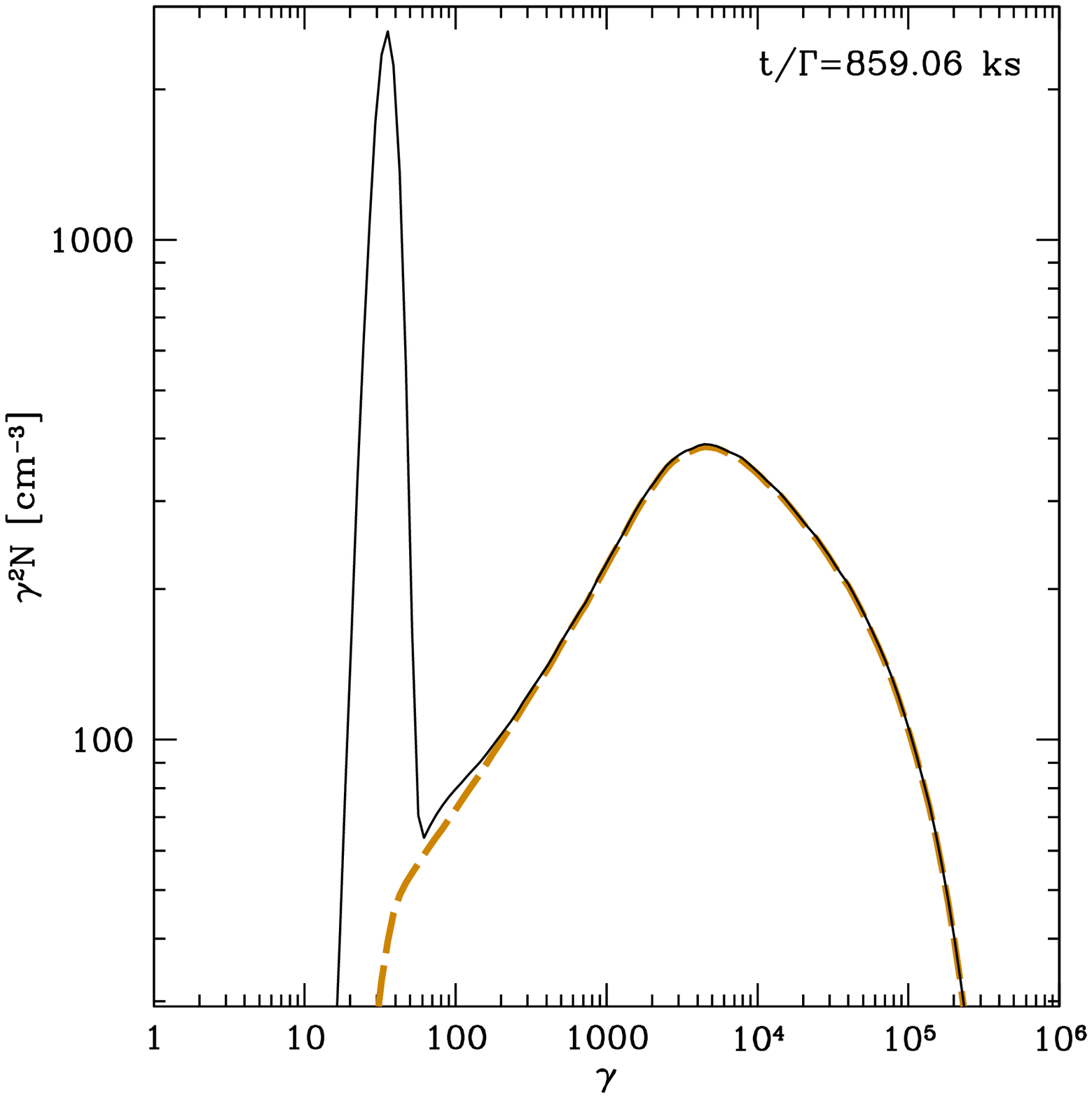}
\includegraphics[width=0.49\linewidth]{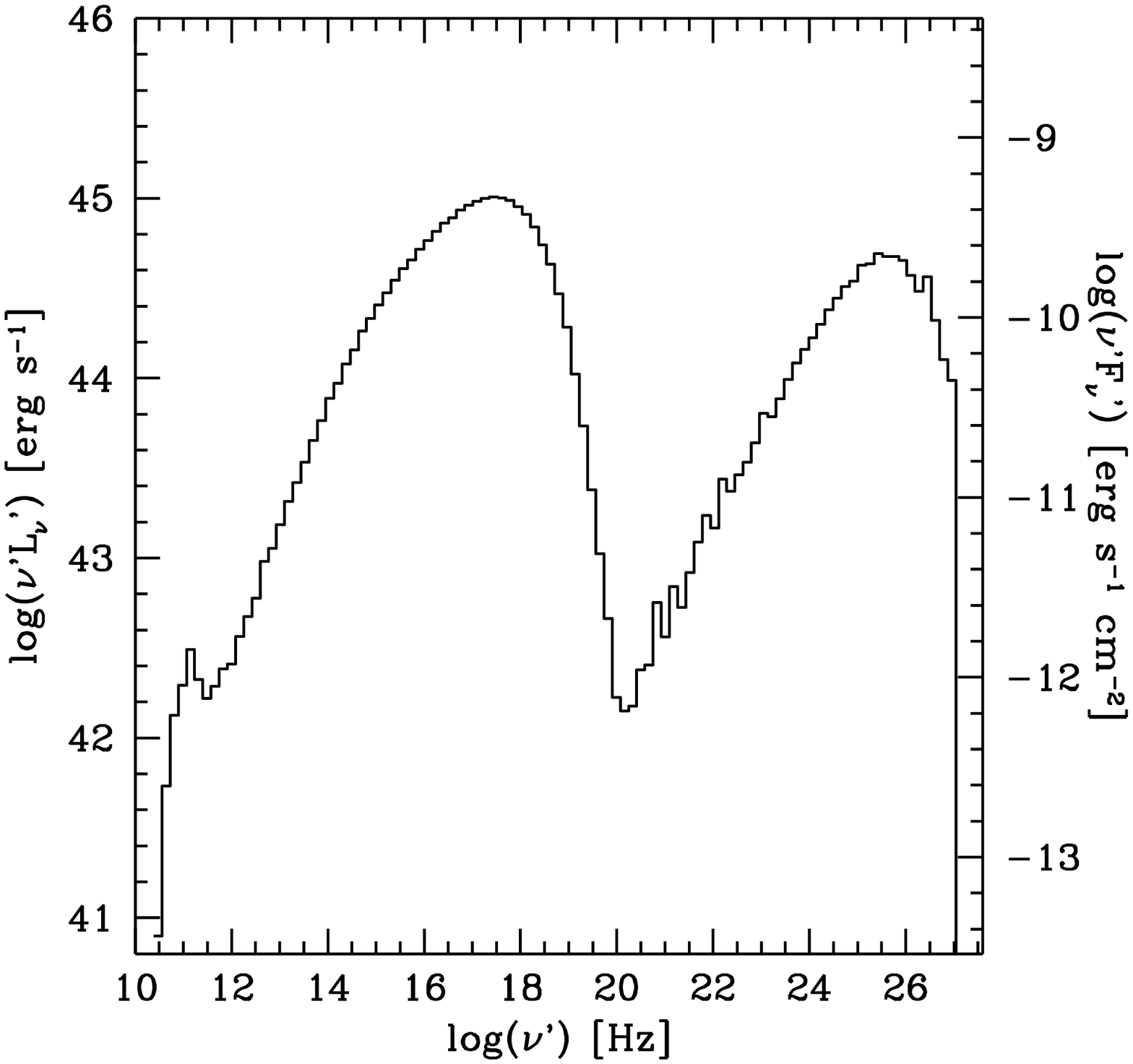}
\caption{Density maps, total EED (black solid line) and SED for the case of \S \ref{esc:away} (open boundary, acceleration away from the center) at simulation time step 1700. The picking up of particles happens at the center of the emission region,
not at the location of the acceleration region, which is what happens in Fig.\,\ref{fig:esc_away}.
The spectral indices of the EED are -1.60 at $\gamma=2\times10^2$ and -2.34 at $\gamma=2\times10^4$.
The EED shown in 
Fig. \ref{fig:esc_away} is plotted here for comparison (orange dashed line).
}
\label{fig:esc_ctaway} 
\end{figure*}

\begin{figure*}
\centering
 \begin{minipage}{0.4\linewidth}
   \includegraphics[width=0.99\linewidth]{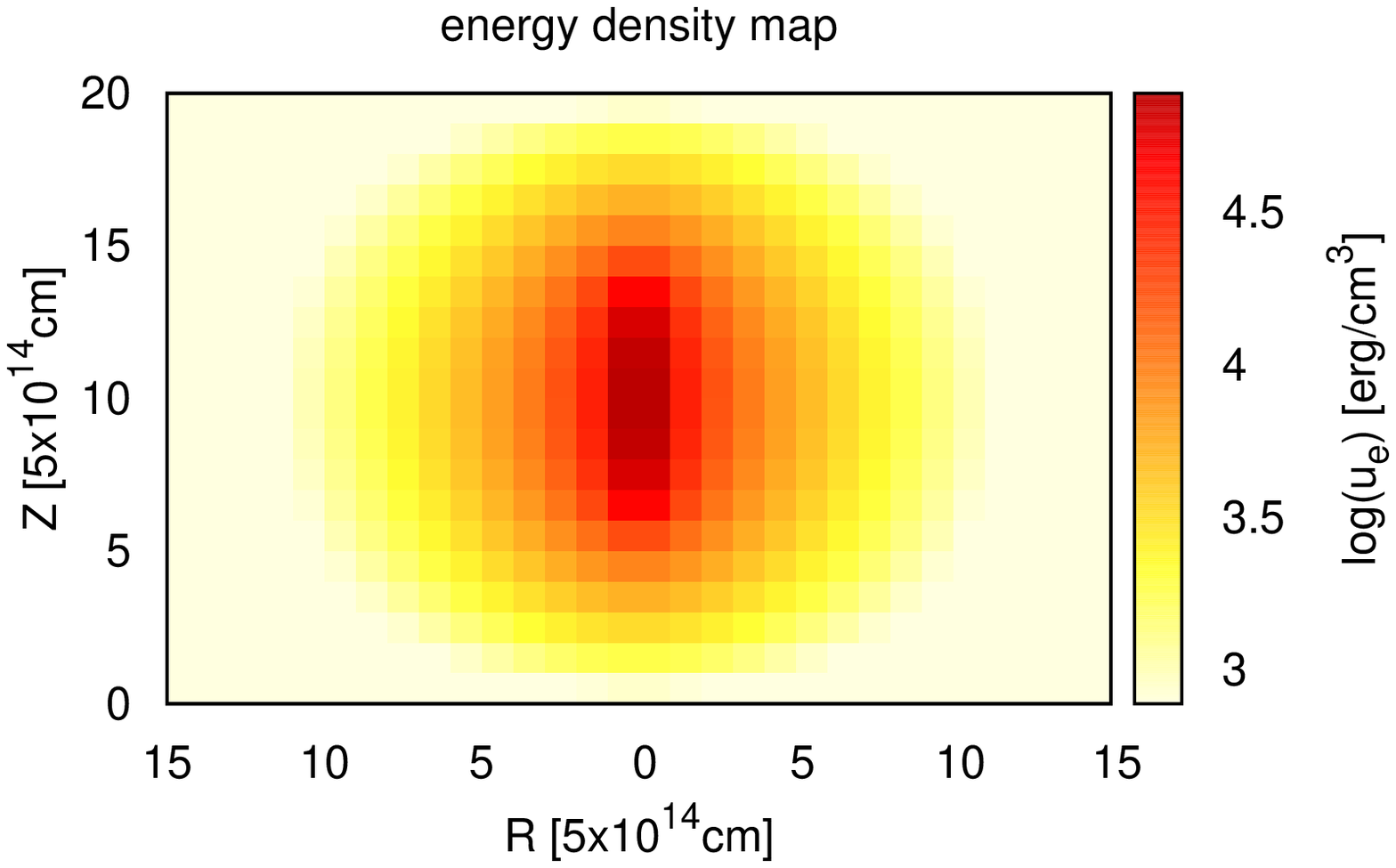}
   \includegraphics[width=0.99\linewidth]{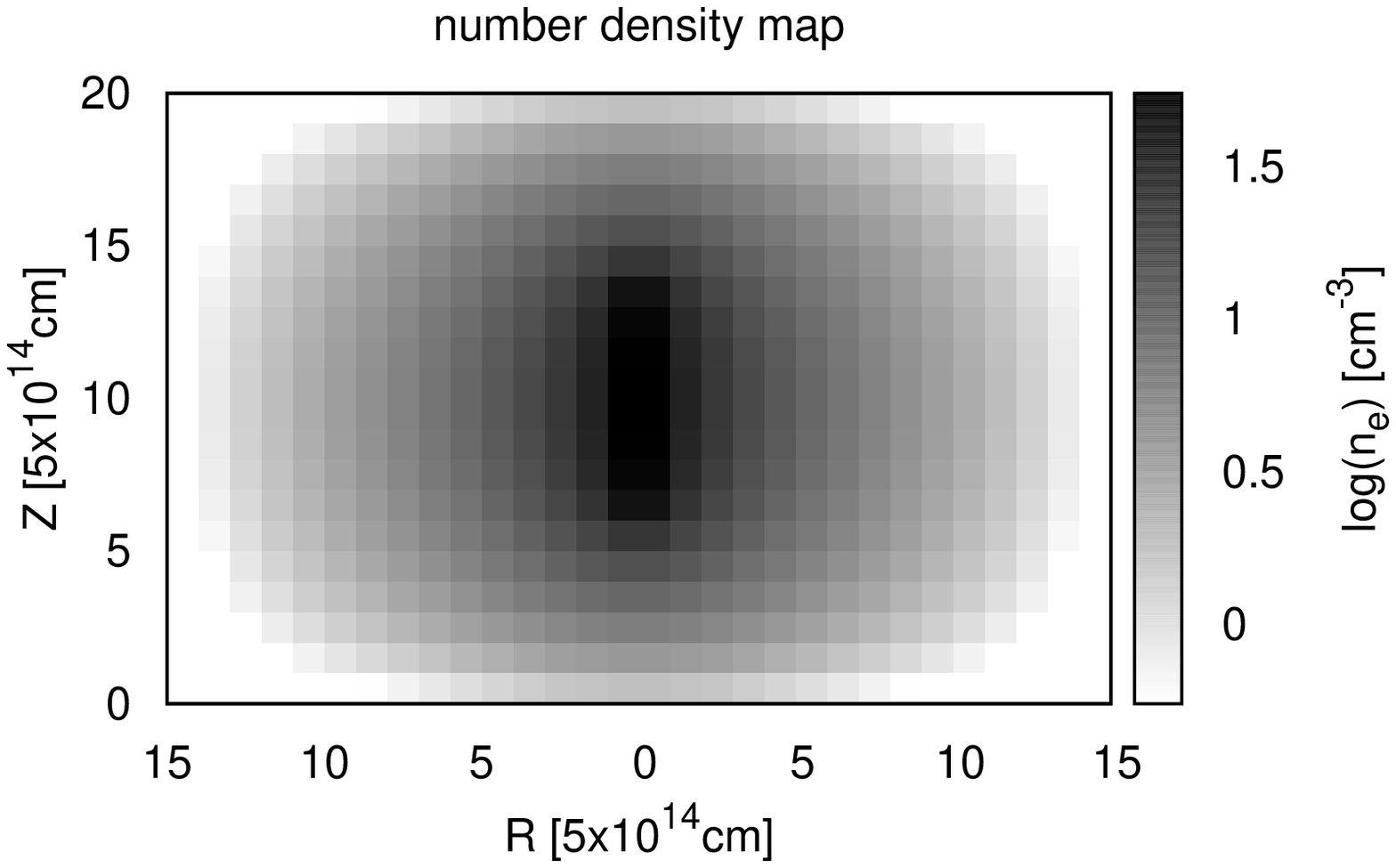}
 \end{minipage}
 \begin{minipage}{0.59\linewidth}
   \includegraphics[width=0.99\linewidth]{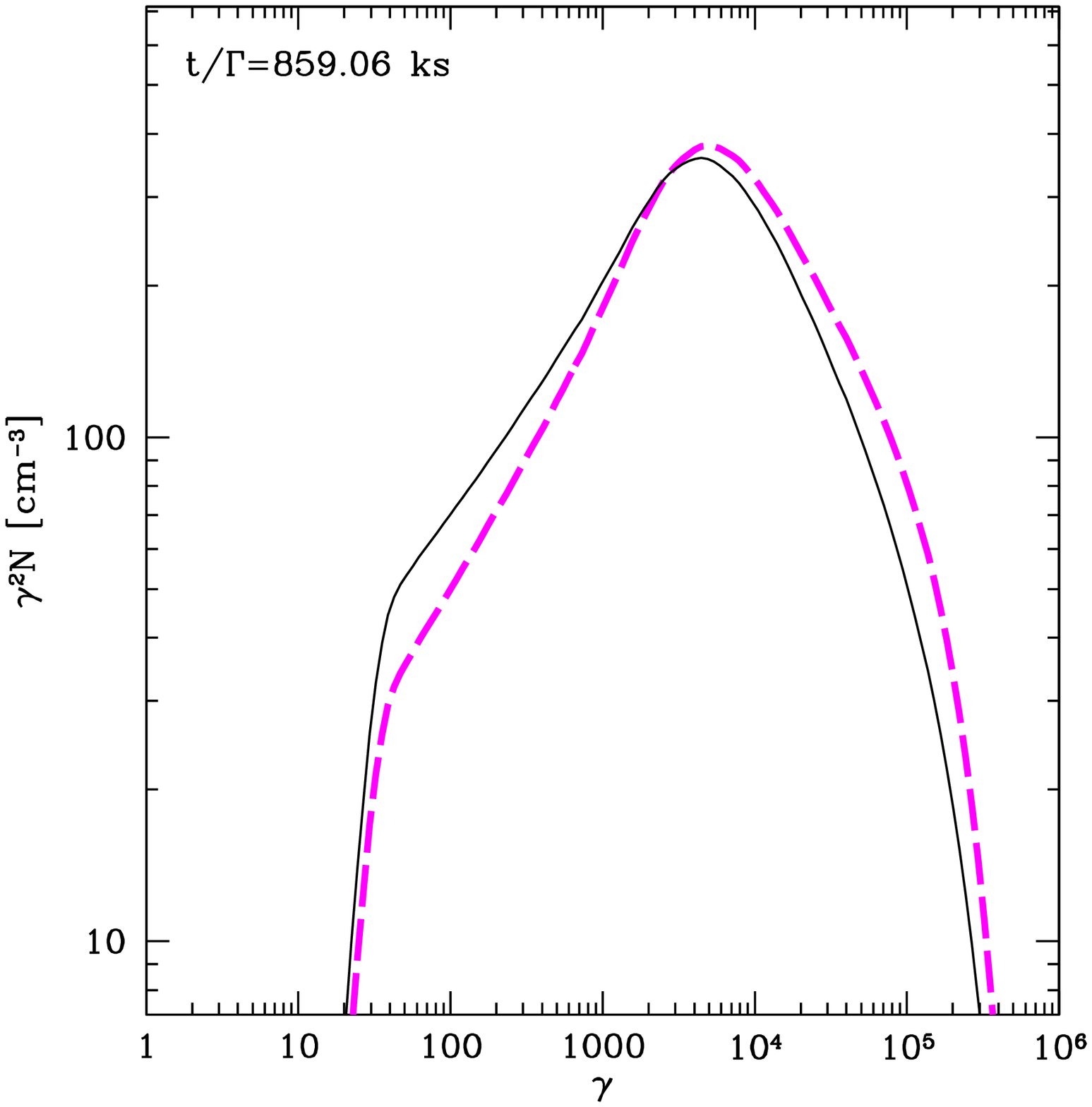}
 \end{minipage}
\caption{Density maps and total EED (black solid line) for the case of 
\S \ref{esc:long} (open boundary, elongated acceleration region) at simulation time step 1700.
The spectral indices of the EED are -1.58 at $\gamma=2\times10^2$ and -2.63 at $\gamma=2\times10^4$.
The final EED shown in
Fig.\,\ref{fig:esc_ct_eed} is plotted here for comparison (magenta dashed line).
}
\label{fig:esc_long} 
\end{figure*}

\subsection{Diffusive Particle escape}
\label{escape}
In this section the boundaries of the emission region are assumed to be open, \ie the particles diffuse outside of the simulation box as if the density outside were zero. 
This implies a constant escape from the emission region,
which is assumed to have a magnetic field stronger than its surroundings so that the emission from the surrounding region is negligible. We also assume that the acceleration
region picks up particles from the intergalactic medium at a constant rate through the turbulent magnetic field.

We have chosen the parameters in the two-zone model (\S \ref{2zone}) so that they are directly comparable 
to the two cases in this section (\S \ref{esc:center} \& \S \ref{esc:slow}). 
However, the total EEDs and SEDs are still slightly different, as can be seen in the comparison in 
Fig.\,\ref{fig:esc_ct_eed} \& \ref{fig:esc_bump}. 
One of the reasons for the difference, for example, is that in the 2D model, the acceleration region 
contains more than one cell. The particle escape time for the central-most cell is longer than 
the escape time for the entire acceleration region, therefore some particles at the highest energy
can have a harder spectrum in the 2D model.
Another example is the posterior consideration of cooling in the acceleration zone of the two-zone model. 
This simplification leads to
the sharp cut-off of EED and SED at the highest energy in the two-zone model, in contrast to the gradual 
cut-off in the 2D model.

\subsubsection{Localized acceleration in the center}
\label{esc:center}
The acceleration region is placed in the center of the emission region, similar to \S \ref{refl:center}.
In addition to the energy density map, we also show the particle number density map in 
Fig.\,\ref{fig:esc_ct_maps}. 
This illustrates how particles are picked up in the central region and then escape from the outer regions.
The total EEDs (Fig.\,\ref{fig:esc_ct_eed} upper-left) at later times overlap with each other, meaning that theyhave already reached a steady state. The steady EED shows a classic broken power-law distribution, with
a cooling break of about 1 at $\gamma=10^3-10^4$. This is consistent with what we saw
in the two-zone model (Fig.\,\ref{fig:2zonect}). The SSC spectrum is also observed to be harder than 
the synchrotron spectrum, a feature already established in the two-zone model, 
and the closed boundary scenario.
In Fig.\,\ref{fig:esc_ct_eed} bottom we plot the EEDs
for three different cells (in the center, mid-way between the center and the outer boundary, and the outer
boundary, all at mid-way height in z direction.)

This case is used as the benchmark case for the open boundary scenario. Main parameters are shown in
Table\,\ref{tab:par}.

\subsubsection{Slow diffusion}
\label{esc:slow}

We study the case with less effecient diffusion, similar to the case in \S\ref{refl:slow}, with the open boundary condition.
In this scenario, we also observe the development of the high-energy bump in the EED,
although it is not obvious in the SED (Fig.\,\ref{fig:esc_bump}).

\subsubsection{Localized acceleration away from the center}
\label{esc:away}

In this section we study the cases where the acceleration region is not located at the center of the emission region. In the first case (Fig.\,\ref{fig:esc_away}),
the acceleration region occupies the 3rd and 4th grid cells from the bottom.
Except the location of the accelerator, the other parameters are identical to those in \S\ref{esc:center}.

Because of the off-center position of the acceleration region, the whole emission region loses particles in
different directions at a different rate. The particle escape happens on several time scales, and can no
longer be described by a single escape time. One consequence of this non-uniform escape is, that the spectral
break in the EED, which is a result of the competition between cooling and escape,
no longer occurs at one specific energy. Instead, the spectrum gradually changes over a large range of
electron energy that likely extends to the cut-off energy. If one were to measure the change in spectral index 
at the break, it would be less than 1, the number expected of radiative cooling.
Because the proximity of the acceleration to the boundary, we also lose particles faster in general. This leads
to a softer `uncooled' spectrum (the one before the break).
The EED with all these effects are shown in Fig.\,\ref{fig:esc_away} bottom, with a comparison to the EED with
acceleration region in the center. An exemplary attempt to measure the spectral change between
$\gamma=2\times10^2$ and $\gamma=2\times10^4 $ gives a break of 0.77.
Compared to \S\ref{esc:center} the average electron density is adjusted to
achieve similar SED and SSC cooling.

In order to test how the proximity of the acceleration to the boundary affects the total EED, we move the
acceleration region closer to the boundary, occupying the 2nd and 3rd grid cells from the bottom
(Fig.\,\ref{fig:esc_away2}).
The EED is shown in comparison with the case above. The measured spectral break becomes even smaller (0.71).
Therefore we predict, if the acceleration region is located further away from the center than our model's
spatial resolution allows, the measured spectral break in the total EED may be significantly smaller than 1.

Another question we address is whether the location of the particle injection 
affects the spectral break. To answer this question we conceive a case where the particles are
injected in the central region with 2x2 cells, while the acceleration region is located
$10^{15}$cm (2 cells) away from the bottom boundary (Fig.\,\ref{fig:esc_ctaway}).
The resulted total EED shows a prominent bump at the injection energy $\gamma=33$. But otherwise the EED
is almost identical to that of Fig.\,\ref{fig:esc_away} (the case with off center injection).
We conclude that the spectral break is not affected by the location of the particle injection,
but only by the location of the particle acceleration.
However, this case is unlikely to represent the real picture of what happens in blazar jets, because it
predicts a flux excess in radio frequency, which is not consistent with observation.

\subsubsection{Elongated acceleration region}
\label{esc:long}

Except the location of the accelerator, we also explore the effect of different geometry of the accelerator.
In this case we construct an elongated accelerator with 8x1 cells (Fig.\,\ref{fig:esc_long} left). 
The total volume of the accelerator is the same as in \S\ref{esc:center}. 
The other parameters are kept unchanged.
The resulting EED (Fig.\,\ref{fig:esc_long} right) has a slightly softer spectrum index both below and above
the spectral break, while the break remains close to 1.
This is caused by the more efficient escape from the accelerator under the current geometry with
unchanged diffusion coefficient. However, without a reference spectrum, 
it is difficult to distinguishes the spectrum from this case from those in \S\ref{esc:center}.
This result indicates that the geometry of the acceleration region has little impact on the EED.
The choice of geometry does not affect our findings regarding the spectral breaks, or spectral hardening
with increasing energy, as discussed in previous sections.

\section{Discussion}
\label{discussion}

\subsection{Spectral hardening at high energy}
\label{dis:hard}
In both the closed and open-boundary scenarios, we notice the spectral hardening of the EED at high energy, if
the particle diffusion is sufficiently slow (\S\ref{refl:slow} \& \S\ref{esc:slow}). This is a result of accounting for acceleration region and
emission region at the same time. The acceleration region, which is small but dominates both the EED and the
SED at high energy, has a harder spectrum compared to the emission region because of radiative cooling.
This dominance in the synchrotron SED will be even stronger if we consider a stronger magnetic field 
in the acceleration region. An exception is that if the magnetic field is so strong that 
the emission from the acceleration region dominates at all energies, the spectral hardening will no longer be
present.
If observations can measure the spectral index accurately enough,
at a frequency close to but below the high energy cut off, 
we could search for this hardening of spectrum.
Its existence will be evidence for localized particle acceleration and moderate particle escape
being at play in AGN jets.
A similar spectral hardening is not clearly visible in the SSC spectra. This might be related to
the broadness of the seed photon spectra in the SSC scenario. Whether the spectral hardening for \grays can be more
apparent in an EC scenario will be assessed in our future work.
Interestingly, at very high energy (VHE, above 100\,GeV) \gray, several blazars are already observed to show hardening of the spectra 
towards higher energy after the correction for extragalactic background light (EBL) absorption 
\citep{albert_etal:2008:3C279_magic_detection, veritas_2014:pks1424:785.16}.

\subsection{Hard SSC spectrum}

In all our results, we observe the SSC spectra in the SED to be significantly harder than the 
synchrotron spectra at corresponding wavelengths. This is caused by the preference of IC scattering
between high-energy synchrotron photons and high-energy electrons, because both of them are concentrated
close to the accelerator in inhomogeneous jet models.
This effect, combined with the hard EED in the cases with slow diffusion,
provides a mechanism to produce very hard spectra (photon index harder than -1.5
\footnote{In a one-zone model, where the acceleration region is usually not accounted for, 
fast synchrotron loss preclude the power-law index of the electron spectra to be harder than -2
regardless of the acceleration mechanism. This implies the photon index can not be
harder than -1.5, under, again, one-zone assumptions \citep{aharonian_2008:hard_spectra.387.1206}.})
at GeV energy at least
(see \S \ref{2zone} and \S \ref{refl:slow}).
Since our choice of parameters is based on the SEDs of \mrk,
a different parameter set might shift those hard spectra to 
even higher energy.
Considering these effects, the inhomogeneous jet model might provide very important explanation for some of 
the unexpectedly hard VHE \gray 
spectra  measured in several `high'-redshift VHE blazars after correction for the 
EBL absorption, and loosen the constraint these observations placed on the EBL \citep{hess_2006:ebl_nature:440.1018}.

\subsection{Electron spectral break}
\label{dis:break}

Radiative cooling normally softens the electron
spectrum by 1. A spectral break is expected to exist at the electron energy where $t_{cool}=t_{esc,em}$. Below this energy
particles do not have enough time to cool before escape from the emission region, while above this energy,
particles becomes softer because of the cooling effect. In the SED this break is expected to be 0.5 
\citep{sari_piran_narayan:1998:spectra_and_lc_of_GRB}.
However, the transition between uncooled and cooled
spectra does not necessarily present itself as a clean cut broken power-law.
In the open boundary scenario, we found (\S\ref{esc:away}) that if the acceleration region is not located in the center of the
emission region, the EED changes gradually over an energy range, and if measured as a broken power-law, 
the break may appear less than 1 (or 0.5 in the SED).
If the observed power-law break in the SED is much larger than 0.5, it can not be explained by the cooling/escape break.

\subsection{Limitation of the simulation}
\label{dis:limit}
Our simulation volume is divided into 20x15 cells. 
Higher-resolution simulations are possible but not practical
because of the computational cost. The acceleration region in our model is therefore set to be of approximately
1/10 the length scale of the emission region, allowing it to occupy 2x2 cells. 
This limit on the acceleration region is only a numerical one, 
but not a physical constraint, \ie the acceleration region can even be smaller, or located closer
to the outer part of the emission region in AGN jets.
Some of the phenomena observed in the current modeling work can be more significant if larger
size ratio is considered.

\section{Conclusions}
\label{conclusion}

We used our inhomogeneous time-dependent emission models to investigate the 
localized particle acceleration and spatial diffusion in AGN jets. This work focus on the steady-state
spectrum and we summarize our findings as follows:

\begin{enumerate}

\item With the acceleration region much smaller than the emission region, the electrons form 
power-law/broken power-law distributions that adequately reproduce blazar SEDs with reasonable rates of
particle escape;
\item The inhomogeneity developed in the jet is energy-dependent, with higher-energy particles concentrated
in smaller regions;
\item The inclusion of particles both inside and outside of the acceleration region causes the EED/SED
to show spectral hardening at high energy, if particle diffusion is slow;
\item The energy-dependent inhomogeneity causes the SSC spectrum to be harder than the synchrotron spectrum,
and this might help to explain the very hard VHE spectra in several blazars.
\item If the acceleration region is not located at the center of the emission region in an open-boundary
 scenario, the resulting EED forms an atypical broken-power-law distribution with spectral break less than 1;
\item The EED formed is weakly dependent on the geometry of the acceleration region.
\end{enumerate}

\section*{Appendix A}
\label{app:a}

Let $N_d(\gamma,t)$ denote the total number spectrum of particles in the diffusion (outer) zone, where escape is possible on time scale $\tau_\mathrm{esc,d}$ and energy loss on time scale $\tau_\mathrm{loss}=1/(a\,\gamma)$, as for a dominance of synchrotron losses. The electron spectrum must satisfy the continuity equation
\be
\frac{\partial N_d(\gamma,t)}{\partial t}-\frac{\partial }{\partial \gamma}\left(\frac{\gamma\,N_d(\gamma,t)}{\tau_\mathrm{loss}}\right)+\frac{N_d(\gamma,t)}{\tau_\mathrm{esc,d}}=Q=\frac{N_a(\gamma,t)}{\tau_\mathrm{esc,a}}\ .
\label{app:1}
\ee
Here, we have already indicated that the source term $Q$ is given by the rate of particle leakage out of the acceleration zone, for which $N_a(\gamma,t)$ denotes the particle spectrum and $\tau_\mathrm{esc,a}$ the escape time scale. The general solution to equation~\ref{app:1} is given by
\begin{align}
N_d(\gamma,t)=& \frac{\tau_\mathrm{loss}}{\gamma} \int_\gamma^\infty dq \int_{-\infty}^t dt'\ \delta\left(t-t'-\frac{1}{a\,\gamma}+\frac{1}{a\,q}\right) \nonumber \\
 &\times\,\frac{N_a(q,t')}{\tau_\mathrm{esc,a}}\,
\exp\left(-\int_\gamma^q \frac{du}{a\,u^2\,\tau_\mathrm{esc,d}}\right)\ .
\label{app:2}
\end{align}
For an energy-independent escape rate we can use the new momentum variable $x=1-\gamma/q$ 
%and w.l.o.g. set $t=0$ 
to simplify equation~\ref{app:2} to
\begin{align}
N_d(\gamma,t)=& \frac{\tau_\mathrm{loss}}{\tau_\mathrm{esc,a}} \int_0^1 dx \int_{-\infty}^0 dt'\ \delta\left(t-t'-x\,\tau_\mathrm{loss}\right) \nonumber \\
&\times\,\frac{N_a \left(\frac{\gamma}{1-x},t'\right)}{(1-x)^2}\,
\exp\left(-x\,\frac{\tau_\mathrm{loss}}{\tau_\mathrm{esc,d}}\right)\ .
\label{app:3}
\end{align}
Note that the exponential is relevant only if the outer boundaries are open, because $\tau_\mathrm{esc,d}=\infty$ and hence $\exp(\dots)=1$ for closed boundaries.

The delta functional in equation~\ref{app:3} is solved by $x=(t-t')/\tau_\mathrm{loss}$, and so we can write
\begin{align}
N_d(\gamma,t)=& \frac{1}{\tau_\mathrm{esc,a}} \int_{t-\tau_\mathrm{loss}}^t dt'\ N_a \left(\frac{\gamma\,\tau_\mathrm{loss}}{\tau_\mathrm{loss}+t'-t},t'\right) \nonumber \\
&\times\,\left(1+\frac{t'-t}{\tau_\mathrm{loss}}\right)^{-2}\,
\exp\left(\frac{t'-t}{\tau_\mathrm{esc,d}}\right)\ .
\label{app:4}
\end{align}

We now need the solution for the electron number spectrum in the acceleration zone, $N_a(\gamma,t)$. It obeys the continuity equation

\begin{align}
\frac{\partial N_a(\gamma,t)}{\partial t}-  \frac{\partial }{\partial \gamma} & \left(\frac{\gamma\,N_a(\gamma,t)}{\tau_\mathrm{loss}} - \frac{\gamma\,N_a(\gamma,t)}{\tau_\mathrm{acc}}+ 
 \frac{\gamma^2}{2\,\tau_\mathrm{acc}}\,\frac{\partial N_a(\gamma,t)}{\partial \gamma}\right) \nonumber \\
& +\frac{N_a(\gamma,t)}{\tau_\mathrm{esc,a}}=Q_a\ .
\label{app:5}
\end{align}
Writing $N_a(\gamma,t)=F_a(\gamma,t)\,\exp\left(-t/\tau_\mathrm{esc,a}\right)$ and assuming that $\tau_\mathrm{loss}\gg \tau_\mathrm{acc}$, i.e. staying away from the loss-induced high-energy cut off in the spectrum, we can simplify equation~\ref{app:5} to
\begin{align}
\frac{\partial F_a(\gamma,t)}{\partial t}+\frac{\partial }{\partial \gamma} & \left(\frac{\gamma\,F_a(\gamma,t)}{\tau_\mathrm{acc}}-\frac{\gamma^2}{2\,\tau_\mathrm{acc}}\,\frac{\partial F_a(\gamma,t)}{\partial \gamma}\right)\nonumber \\
& =Q_a\,\exp\left(\frac{t}{\tau_\mathrm{esc,a}}\right)\ .
\label{app:6}
\end{align}
Under the condition $\tau_\mathrm{acc}\neq \tau_\mathrm{acc}(\gamma)$ we can find Green's function for this problem in the literature \citep{kardashev:1962}. With the source term on the right-hand side, the solution of equation~\ref{app:6} is
\begin{align}
N_a(\gamma,t)&= \frac{\sqrt{\tau_\mathrm{acc}}}{\sqrt{2\,\pi}\,\gamma}\,\int_1^\infty d\gamma'\ \int_0^t dt'\ 
\frac{Q_a(\gamma',t')}{\sqrt{t-t'}}\,\exp\left(-\frac{t-t'}{\tau_\mathrm{esc,a}}\right) \nonumber \\
&\times\,\exp\left(-\left[\sqrt{\frac{\tau_\mathrm{acc}}{2\,(t-t')}}\,\ln \frac{\gamma'}{\gamma} +\frac{3}{2}\,\sqrt{\frac{(t-t')}{2\,\tau_\mathrm{acc}}}
\right]^2\right)\ .
\label{app:7}
\end{align}
The differential number density of electron in the diffusion zone would be given by $n_d(\gamma,t) =N_d(\gamma,t) / V_d$, where $V_d$ is the volume of the diffusion zone. Likewise, $n_a(\gamma,t) =N_a(\gamma,t) / V_a$ is the differential density in the acceleration zone. When $\tau_\mathrm{esc,d}$ is finite, the rate of transfer back into the acceleration zone is a small fraction of the total escape rate from the diffusion zone, which in the steady state itself is a lower limit to the escape rate from the acceleration zone. 
Therefore return flux is not an issue in open boundary
situations modeled in a two-zone approach. 
In a closed boundary scenario, however, the total amount of escape and return particles are equal,
and the return flux should be important.
\section*{Acknowledgements}

We thank the anonymous referee for constructive suggestions that helped improving the paper.
XC and MP acknowledge support by the Helmholtz Alliance for Astroparticle Physics HAP funded by the Initiative and Networking Fund of the Helmholtz Association. MB acknowledges support from the South African Department of Science and Technology through the National Research Foundation under NRF SARChI Chair grant No. 64789.

\bibliography{refs_all}

\end{document}